\documentclass[aps,prd,amssymb,nofootinbib,twocolumn,epsf,floatfix]{revtex4}
\usepackage[usenames]{color}
\usepackage[normalem]{ulem} 
\usepackage{graphicx} 
\usepackage{subfigure} 
\usepackage{amssymb}  
\usepackage{amsmath}
\usepackage{mathrsfs}
\usepackage{epstopdf}
\usepackage{float}
\usepackage{booktabs} % for tabular with \toprule, \midrule, \bottomrule
\usepackage{microtype}
\makeatletter
\AtBeginDocument{% 
  \def\@makecaption#1#2{%
    \par
    \vskip\abovecaptionskip
    \begingroup 
     \footnotesize\rmfamily
     \sbox\@tempboxa{%
      \let\\\heading@cr
      #1: #2%
     }%
     \@ifdim{\wd\@tempboxa >\hsize}{%
      \begingroup
       \samepage
       \flushing
       \let\footnote\@footnotemark@gobble
       #1: #2\par
      \endgroup
     }{%
       \global \@minipagefalse
       \hb@xt@\hsize{\hfil\unhbox\@tempboxa\hfil}%
     }%
    \endgroup
    \vskip\belowcaptionskip
  }%
}
\makeatother
\usepackage[colorlinks,breaklinks]{hyperref}

\newcommand{\llangle}{\langle\kern-0.2em \langle}
\newcommand{\rrangle}{\rangle\kern-0.2em \rangle}

\begin{document}
 
%%%%%%%%%%%%%%%%%%%%%%%%%%%%%%%%%%%%%%%%%%%%%%%%%%%%%%%%%%%%%%%%%%%%%%%%%%%%%%%
% Constructing Homogeneous Cosmological Models Numerically Using Ricci Flow
%%%%%%%%%%%%%%%%%%%%%%%%%%%%%%%%%%%%%%%%%%%%%%%%%%%%%%%%%%%%%%%%%%%%%%%%%%%%%%%
\title{Using Ricci Flow in the Construction of Homogeneous
  Cosmological Models}

\author{Fan Zhang\,${}^{a,b}$ and Lee Lindblom\,${}^c$}

\affiliation{${}^a$Institute for Frontiers in Astronomy and Astrophysics,
  Beijing Normal University, Beijing 102206, China}

\affiliation{${}^b$Gravitational Wave and Cosmology Laboratory,
  School of Physics and Astronomy,
  Beijing Normal University, Beijing 100875, China}

\affiliation{${}^c$ Department of Physics, University 
  of California at San Diego, La Jolla, CA 92093, USA}

\date{\today} 

\begin{abstract}
  This study develops numerical methods for computing homogeneous and
  isotropic cosmological models on manifolds with non-trivial spatial
  topologies.  The use of Ricci flow is introduced here to aid in the
  construction of globally homogeneous initial data on these
  manifolds.  The initial data produced in this way are evolved by
  solving Einstein's equation numerically to construct homogeneous and
  isotropic cosmological models on three exemplar manifolds in this
  study: Seifert-Weber space, the lens space $L(5,2)$, and Sixth-Turn
  space.
\end{abstract}
 
\maketitle

%%%%%%%%%%%%%%%%%%%%%%%%%%%%%%%%%%%%%%%%%%%%%%%%%%%%%%%%%%%%%%%%%%%%%%%%%%%%%%%
% Introduction
%%%%%%%%%%%%%%%%%%%%%%%%%%%%%%%%%%%%%%%%%%%%%%%%%%%%%%%%%%%%%%%%%%%%%%%%%%%%%%%
\section{Introduction}
\label{s:Introduction}

The topology of our universe is presently unknown.  Several attempts
to observe topological imprints in the cosmic microwave radiation have
been made, but no definitive topology related structure has yet been
detected, see Refs.~\cite{Copi2026,LachiezeRey1995}.  Theoretical
analyses of possible observable topological effects have been limited
to simple cases where the structures of those spaces are known
analytically, e.g. see Refs.~\cite{LachiezeRey1995, riazuelo2004,
  Peng2019}. The purpose of this study is to expand the set of
computational tools that facilitate the exploration of cosmological
models numerically on a much larger collection of manifolds with
non-trivial spatial topologies.

The large-scale geometry of our universe is observed to be nearly
homogeneous and isotropic, see Refs.~\cite{Planck2015XIII,
  Planck2015XVI}. Modern high precision cosmological observations have
discovered a few anomalies, like the Hubble tension~\cite{Cai2026} and
the dipole anomaly~\cite{Secrest2025}, that are not accounted for in
the standard Friedman-Robertson-Walker models.  But these anomalies
represent relatively small deviations from the standard models,
therefore realistic cosmological models on manifolds with any topology
must be nearly homogeneous and isotropic. A previous study,
Ref.~\cite{Zhang2026a}, showed that it was not possible to construct
homogeneous initial data on many manifolds simply by solving a simple
conformal version of the Einstein initial value equations.  This study
introduces the use of Ricci flow as a tool in the construction of
realistic cosmological initial data that can overcome that limitation
in many cases.  Ricci flow is a parabolic equation for the spatial
metric that drives the geometry toward a uniform curvature state, see
Refs.~\cite{Hamilton1982, Chow2004}.  This study demonstrates this use
of Ricci flow by constructing homogeneous and isotropic initial data
on manifolds with a range of non-trivial topologies.

The numerical methods used in this study are based on a series of
papers, Refs.~\cite{Lindblom2013, Lindblom2014, Lindblom2015,
  Lindblom2022, Zhang2022, Lindblom2024, Zhang2026a}, that develop
methods for solving partial differential equations on manifolds with
arbitrary spatial topologies.  These methods were created to solve
Einstein's evolution equation on manifolds with topology
$\mathbb{R}\times\Sigma$.  The spacelike surfaces, $\Sigma$, are
represented in these numerical methods as multi-cube structures that
consist of collections of non-overlapping cubic coordinate charts
whose faces are identified with the faces of neighboring cubes by a
collection of maps, see Ref.~\cite{Lindblom2013}.  These maps define
the topological structure of $\Sigma$ that determines the continuity
of scalar fields across the multi-cube interfaces. The differential
structures on these multi-cube representations (i.e.  the structures
that define the continuity of vectors and tensors across the
multi-cube interfaces) are fixed by a $\mathcal{C}^1$ reference metric
$\tilde g_{ij}$ that is specified as part of the multi-cube structure.
Suitable reference metrics have been constructed numerically for the
manifolds included in this study using the methods described in
Refs.~\cite{Lindblom2022, Lindblom2024}.  The numerical
methods used in this study have been used previously to solve
Einstein's equation on a small collection of manifolds, see
Refs.~\cite{Lindblom2014, Zhang2026a, Zhang2026b}. These numerical
methods have also been used to solve the Ricci flow equation on
two-dimensional manifolds with non-trivial topologies, see
Ref.~\cite{Lindblom2015}.  Those methods are extended here to provide
the numerical solutions to the Ricci flow equation used in the
construction of the initial data for Einstein's equation on spatial
manifolds, $\Sigma$, from a large collection of three-dimensional
manifolds.

This study has constructed numerical solutions to the Ricci flow
equation, and from them homogeneous and isotropic cosmological models,
on three exemplar manifolds: Seifert-Weber space, the lens space
$L(5,2)$, and Sixth-Turn space.  Seifert-Weber space is constructed
topologically by identifying opposite faces of a dodecahedron after
twisting the faces by the angle $3\pi/5$, see
Ref.~\cite{SeifertWeber1933}.  It belongs to the hyperbolic geometry
class of manifolds that admit negative scalar curvature geometries.
The lens space $L(5,2)$ is constructed by identifying points on the
three-sphere $S^3$ which are related by a discrete isometry specified
by the integers $p$ and $q$.  In particular, let the complex numbers
$(z_1,z_2)$ represent a point on $S^3$ as a point on the unit sphere
in $\mathbb{C}^2$.  The lens space $L(p,q)$ is the quotient space that
identifies the point $(z_1,z_2)$ on $S^3$ with the point $(e^{2\pi
  i/p}\, z_1, e^{2\pi i q/p}\,z_2)$, see Refs.~\cite{ThrelfallSeifert1931,
  Wolf2011}.  These lens spaces belong to the $S^3$ geometry class of
manifolds which admit positive scalar curvature geometries.  The
Sixth-Turn space can be constructed topologically by identifying
opposite rectangular faces of a hexagonal cylinder, and then
identifying the two hexagonal faces after twisting by $2\pi/6$, see
Ref.~\cite{Wolf2011}.  The Sixth-Turn space (also called $E5$ in some
literature, see Ref.~\cite{Scott1983, Wolf2011}) is one of the six
compact orientable three-manifolds that admits a global flat metric.
The multi-cube structures used in this study for these manifolds are
given explicitly as tables in Ref.~\cite{Lindblom2022}: Table D.7 for
the Seifert-Weber space, Table D.6 for the lens space $L(5,2)$, and
Table D.2 for Sixth-Turn space.

This paper is organized as follows: Section~\ref{s:RicciFlowEquations}
describes the volume preserving form of the three-dimensional Ricci
flow equation used in this study.  This section also describes a
collection of norms used to measure the homogeneity and isotropy of
the geometries produced by the Ricci flow evolutions.

Section~\ref{s:RicciFlowNumericalResults} describes the numerical
Ricci flow evolutions performed for this study on Seifert-Weber space,
the lens space $L(5,2)$, and the Sixth-Turn space.  These numerical
results are shown to be convergent as $N_\mathrm{grid}$ (the number of
grid points in each direction in each multi-cube region) increases.
In particular, these results are shown to converge toward homogeneous
and isotropic geometries as $N_\mathrm{grid}$ increases.

Section~\ref{s:InitialData} uses the results of these Ricci flow
evolutions as the conformal metrics in the solution of the Einstein
initial value problem.  The initial value problem used here is a
particularly simple one in which the cosmological constant is assumed
to dominate over the normal matter content of the universe, see
Ref.~\cite{Zhang2022}.

Section~\ref{s:CosmologicalModels} evolves
these initial data using a covariant symmetric-hyperbolic
representation of Einstein's equation, which is summarized in some
detail in Appendix~\ref{s:SymmetricHyperbolicSystem}.  These
evolutions are shown to produce spacetimes that converge toward simple
homogeneous and isotropic cosmological models.  Although expressed in
very different coordinates, these numerical cosmological models are
shown to have the same expansion histories as the analogous simple
Friedman-Robertson-Walker cosmologies.

Finally, Sec.~\ref{s:Discussion} summarizes and discusses the new
results of this study, describes some limitations to these results,
and presents some ideas for future studies.

%%%%%%%%%%%%%%%%%%%%%%%%%%%%%%%%%%%%%%%%%%%%%%%%%%%%%%%%%%%%%%%%%%%%%%%%%%%%%%%
\section{Ricci Flow: Equations}
\label{s:RicciFlowEquations}
%%%%%%%%%%%%%%%%%%%%%%%%%%%%%%%%%%%%%%%%%%%%%%%%%%%%%%%%%%%%%%%%%%%%%%%%%%%%%%%

This section presents the version of the three-dimensional Ricci flow
equation used in this study.  This section also describes the
numerical method used to enforce the appropriate boundary conditions
at the interfaces between multi-cube coordinate patches.  Several
norms are also defined in this section that can be used to measure the
homogeneity and isotropy of the spatial geometries produced by the
Ricci flow evolutions.

Ricci flow is a parabolic evolution equation for
the spatial metric $g_{ij}$, which can be written as
\begin{eqnarray}
\partial_t g_{ij} &=& F_{ij},
\label{e:RicciFlow}
\end{eqnarray}
where $F_{ij}$ is given by
\begin{eqnarray}
  F_{ij} &=& - 2 R_{ij}+ \nabla_i \Delta_j + \nabla_j \Delta_i
  \nonumber\\ && +\tfrac{2}{3}\,\llangle R(t)\rrangle\,g_{ij}
  -\tfrac{2}{3}\,\mu\frac{\mathcal{V}(t)-\mathcal{V}_0}{\mathcal{V}(t)}\,g_{ij},
\label{e:RicciFlowRHS}
\end{eqnarray}
see Ref.~\cite{Hamilton1982}.  The principal terms on the right side
of Eq.~(\ref{e:RicciFlowRHS}) are the Ricci tensor, $R_{ij}$, and the
terms containing the covector, $\Delta_i$, defined by
\begin{eqnarray}
  \Delta_i=g_{ij}g^{k\ell}(\Gamma^j_{k\ell}-\tilde \Gamma^j_{k\ell}).
  \label{e:DeltaDef}
\end{eqnarray}
The connection $\Gamma^j_{k\ell}$ that appears in
Eq.~(\ref{e:DeltaDef}), is the Christoffel connection associated with
the metric $g_{ij}$, while $\tilde \Gamma^j_{k\ell}$ is any other
fixed connection on the manifold.  These ``DeTurck'' terms (those
containing $\Delta_i$) are added to make Eq.~(\ref{e:RicciFlow})
strongly parabolic, and thus to have a well-posed initial value
problem, see Refs.~\cite{DeTurck1983, Garfinkle2008}.

The terms in Eq.~(\ref{e:RicciFlowRHS}) containing the spatially
averaged Ricci scalar curvature, $\llangle R\,\rrangle$,
\begin{eqnarray}
\llangle R\, \rrangle = \frac{\int R \sqrt{g}\, d^{\,3}x}{\mathcal{V}},
\label{e:RAverage}
\end{eqnarray}
and the spatial volume $\mathcal{V}$,
\begin{eqnarray}
\mathcal{V} = \int\!\sqrt{g}\, d^{\,3}x.
\label{e:Volume}
\end{eqnarray}
have been added to Eq.~(\ref{e:RicciFlow}) to control the evolution of
the volume of the  manifold.  Contracting Eq.~(\ref{e:RicciFlow}) with
the inverse metric $g^{ij}$ and integrating the result over any
compact manifold without boundary provides the following evolution
equation for the volume $\mathcal{V}(t)$ of that manifold:
\begin{eqnarray}
  \partial_t\left[\mathcal{V}(t)-\mathcal{V}_0\right]=
  -\mu\left[\mathcal{V}(t)-\mathcal{V}_0\right].
\end{eqnarray}
Without the term proportional to $\mu$, the volume of the manifold
would be fixed, $\partial_t \mathcal{V}(t) = 0$, at the analytical
level.  In numerical simulations, however, discretization and roundoff
error give rise to slow, approximately linear drifts in the volume.
With the damping term, however, the volume of the manifold is driven
toward the target value $\mathcal{V}_0$ at a rate determined by the
constant $\mu$.  The numerical value $\mu=10$ was found to be useful
in the evolutions for this study.

Penalty terms have been added to the Ricci flow equations at the
boundaries between multi-cube regions to ensure the metric and its
derivatives are continuous across those boundaries.  At these boundary
interface points the standard Ricci flow equation,
Eq.~(\ref{e:RicciFlow}), is replaced with
\begin{eqnarray}
  \partial_t g_{ij} &=& F_{ij} - \alpha\,\Bigl( g_{ij}
  - \langle g_{ij}\rangle\Bigr)\nonumber\\
  &&- \beta\,\left(\tilde n^k\tilde\nabla_k\,g_{ij}
  + \left\langle \tilde n^k\tilde\nabla_k\,g_{ij}\right\rangle\right).
  \label{e:RicciFlowBC}
\end{eqnarray}
The vector $\tilde n^k$ is the unit normal to the boundary surface
using the reference metric $\tilde g_{ij}$ that defines the
differential structure of the manifold: $\tilde n^m\tilde n^n\,\tilde
g_{mn}=1$.  The tensor $\tilde n^k\tilde\nabla_k\,g_{ij}$ is the
normal component of the torsion-free reference-metric compatible
covariant derivative of $g_{ij}$:
\begin{equation}
  \tilde n^k\tilde\nabla_k g_{ij} = \tilde n^k\partial_k g_{ij}
  - \tilde n^k\tilde \Gamma^l_{ki}\, g_{lj}
  - \tilde n^k\tilde \Gamma^l_{kj}\, g_{il}.
  \label{e:NormalCovDerivBC}
\end{equation}
The expressions $\langle g_{ij}\rangle$ and $\langle \tilde
n^k\tilde\nabla_k\,g_{ij}\rangle$ represent the values of those
tensors on the boundaries of the neighboring multi-cube region that
have been transformed into the local coordinate basis using the
boundary interface maps (see Ref.~\cite{Lindblom2014} for details).
The constants $\alpha$ and $\beta$ determine the sizes of these
penalty terms.  The values of these constants used in the evolutions
for this study are
\begin{eqnarray}
  \alpha &=& N_\mathrm{grid}^2,\\
  \beta &=& 1, 
\end{eqnarray}
where $N_\mathrm{grid}$ is the number of grid points in the normal
direction of the particular region.  These values of $\alpha$ and
$\beta$ ensure the boundary conditions $g_{ij} = \langle
g_{ij}\rangle$ and $\tilde n^k\tilde\nabla_k\,g_{ij} =\langle \tilde
n^k\tilde\nabla_k\,g_{ij}\rangle$ are accurately enforced without
making the evolution numerically unstable.
Equation~(\ref{e:RicciFlowBC}) therefore enforces continuity across
the boundary of both the metric and its normal derivatives.

Several norms have been introduced to measure how well the numerical
Ricci flow evolutions are driven toward a homogeneous and isotropic
state.  The first norm measures the spatial variations in the Ricci
scalar curvature, $R$.  The average value of these spatial variations
is given by $\llangle \Delta R\,\rrangle$, defined by
\begin{eqnarray}
\llangle\,\Delta R\,\rrangle^2 = 
\frac{1}{\mathcal{V}}
\int \bigl(R - \llangle R \,\rrangle\bigr)^2 \sqrt{g}\, d^{\,3}x.
     \label{e:RFluctuation}
\end{eqnarray}
A dimensionless norm, $\|\,\Delta R\,\|$, defined by
\begin{eqnarray}
\|\,\Delta R\,\| = 
\frac{\llangle\,\Delta R\,\rrangle}
     {|\llangle R\,\rrangle|},
     \label{e:RFluctuationNorm}
\end{eqnarray}
can also be useful except in cases where $\llangle R\,\rrangle\approx
0$.

The homogeneity of the full spatial Ricci tensor can be explored
further using the curvature scalar $\mathscr{R}$ defined by
\begin{equation}
  \mathscr{R}^2 = 3 R_{ij}R^{ij}.
  \label{e:CurvatureScalarDef}
\end{equation}
Its spatial average $\llangle\mathscr{R}\,\rrangle$ is  defined by
\begin{eqnarray}
  \llangle\mathscr{R}\,\rrangle
  &=& \frac{1}{\mathcal{V}}\int \mathscr{R}\sqrt{g}\,d^{\,3}x,
  \label{e:CurvatureScalarAve}
\end{eqnarray}
and its average spatial variations
$\llangle\,\Delta\mathscr{R}\,\rrangle$ are given by,
\begin{equation}
  \llangle\,\Delta\mathscr{R}\,\rrangle^2
  = \frac{1}{\mathcal{V}}\int \bigl(\mathscr{R} 
    -\llangle\mathscr{R}\,\rrangle\bigr)^2\sqrt{g}\, d^{\,3}x.
  \label{e:CurvatureScaleFluct}
\end{equation}
A dimensionless norm $\|\,\Delta\mathscr{R}\,\|$ defined by
\begin{equation}
  \|\,\Delta\mathscr{R}\,\|
  = \frac{ \llangle\,\Delta\mathscr{R}\,\rrangle}
  {\mathcal{\llangle\mathscr{R}\,\rrangle}},
  \label{e:CurvatureScaleFluctNorm0}
\end{equation}
can also be useful except in cases where $\llangle
\mathscr{R}\,\rrangle\approx 0$.

If the Ricci tensor is determined completely by the Ricci scalar,
$R_{ij}=\tfrac{1}{3}g_{ij}R$, then its eigenvalues are identical and
the geometry is isotropic.  In this case the curvature scalar
$\mathscr{R}$ equals the Ricci scalar: $\mathscr{R}^{\,2}=R^2$.
Conversely if $\mathscr{R}^{\,2}=R^2$ then the spatial geometry must
be isotropic.  Consider a frame in which the Ricci tensor is diagonal,
with eigenvalues $R_1$, $R_2$ and $R_3$.  The Ricci scalar $R$ is
given in this case by $R=R_1+R_2+R_3$, while the curvature scalar
$\mathscr{R}$ is given by $\mathscr{R}^{\,2}=3(R_1^2+R_2^2+R_3^2)$.
Using these expressions it follows that
\begin{equation}
  \mathscr{R}^{\,2}-R^2=(R_1-R_2)^2 + (R_1-R_3)^2 + (R_2-R_3)^2.
\end{equation}
Therefore, if $\mathscr{R}^{\,2}=R^2$ then $R_1=R_2=R_3$, so the geometry
is isotropic.  One practical measure of the isotropy of the geometry
is given by the quantity $\llangle \mathscr{I}\rrangle$ defined by
\begin{equation}
  \llangle\mathscr{I}\rrangle
  =\bigl| |\llangle R\,\rrangle | -\llangle\mathscr{R}\,\rrangle\bigr|.
  \label{e:IsotropyMeasure}
\end{equation}
The dimensionless
norm $\|\mathscr{I}\|$ defined by
\begin{equation}
  \|\mathscr{I}\|
  =\frac{  \llangle\mathscr{I}\rrangle}
  {|\llangle R\,\rrangle|},
  \label{e:IsotropyNorm}
\end{equation}
can also be useful except in cases where $\llangle R\,\rrangle\approx
0$.

%%%%%%%%%%%%%%%%%%%%%%%%%%%%%%%%%%%%%%%%%%%%%%%%%%%%%%%%%%%%%%%%%%%%%%%%%%%%%%%
\section{Ricci Flow: Numerical Results}
\label{s:RicciFlowNumericalResults}
%%%%%%%%%%%%%%%%%%%%%%%%%%%%%%%%%%%%%%%%%%%%%%%%%%%%%%%%%%%%%%%%%%%%%%%%%%%%%%%

This section describes the numerical methods used in this study to
solve the Ricci flow equation on three-dimensional manifolds with
non-trivial topologies.  The geometries produced by these Ricci flow
solutions will be shown to be homogeneous and isotropic.  These
geometries do not satisfy Einstein's equation, but they can be used to
construct simple homogeneous and isotropic cosmological models that do
satisfy Einstein's equation in Secs.~\ref{s:InitialData} and
\ref{s:CosmologicalModels}.

The volume-normalized Ricci flow equation with DeTurck gauge fixing,
Eq.~(\ref{e:RicciFlow}), has been implemented for this study in the
\verb!SpEC!  numerical relativity code (developed originally by the
Caltech and Cornell numerical relativity groups).  This code evolves
PDEs using pseudo-spectral methods to evaluate spatial derivatives,
and performs explicit time integration at each grid point using
standard ordinary differential equation solvers (e.g., the
Dormand-Prince, Refs.~\cite{Dormand1987, Dormand1996}, solver used in
this study).  The Ricci flow equation is parabolic and has been
implemented in \verb!SpEC! as a FOTSOS
(First-Order-Time-Second-Order-Space) evolution system.  Boundary
conditions are imposed at multi-cube interface boundaries to enforce
continuity of the metric $g_{ij}$ and the normal covariant derivative
$\tilde n^k\tilde\nabla_k g_{ij}$ using penalty methods as described
in Sec.~\ref{s:RicciFlowEquations}, and in more detail in
Ref.~\cite{Lindblom2022}.  In this study the reference connection
$\tilde \Gamma^k_{ij}$ that appears in the DeTurck terms of the Ricci
flow Eq.~(\ref{e:RicciFlow}) is the one associated with the reference
metric $\tilde g_{ij}$ constructed numerically for each manifold using
the methods described in Ref.~\cite{Lindblom2022}.\footnote{ Reference
metrics were constructed on each multi-cube manifold by first
constructing flat metrics in the neighborhood of each multi-cube
vertex, then combining those flat metrics with a partition of unity to
form a global $\mathcal{C}^0$ metric, and finally smoothing those
global $\mathcal{C}^0$ metrics across each interface between cubic
regions to create a global $\mathcal{C}^1$ metric by solving a
sequence of differential equations numerically.}

\begin{figure}[!b]
{\centering
\includegraphics[height=0.32\textwidth]{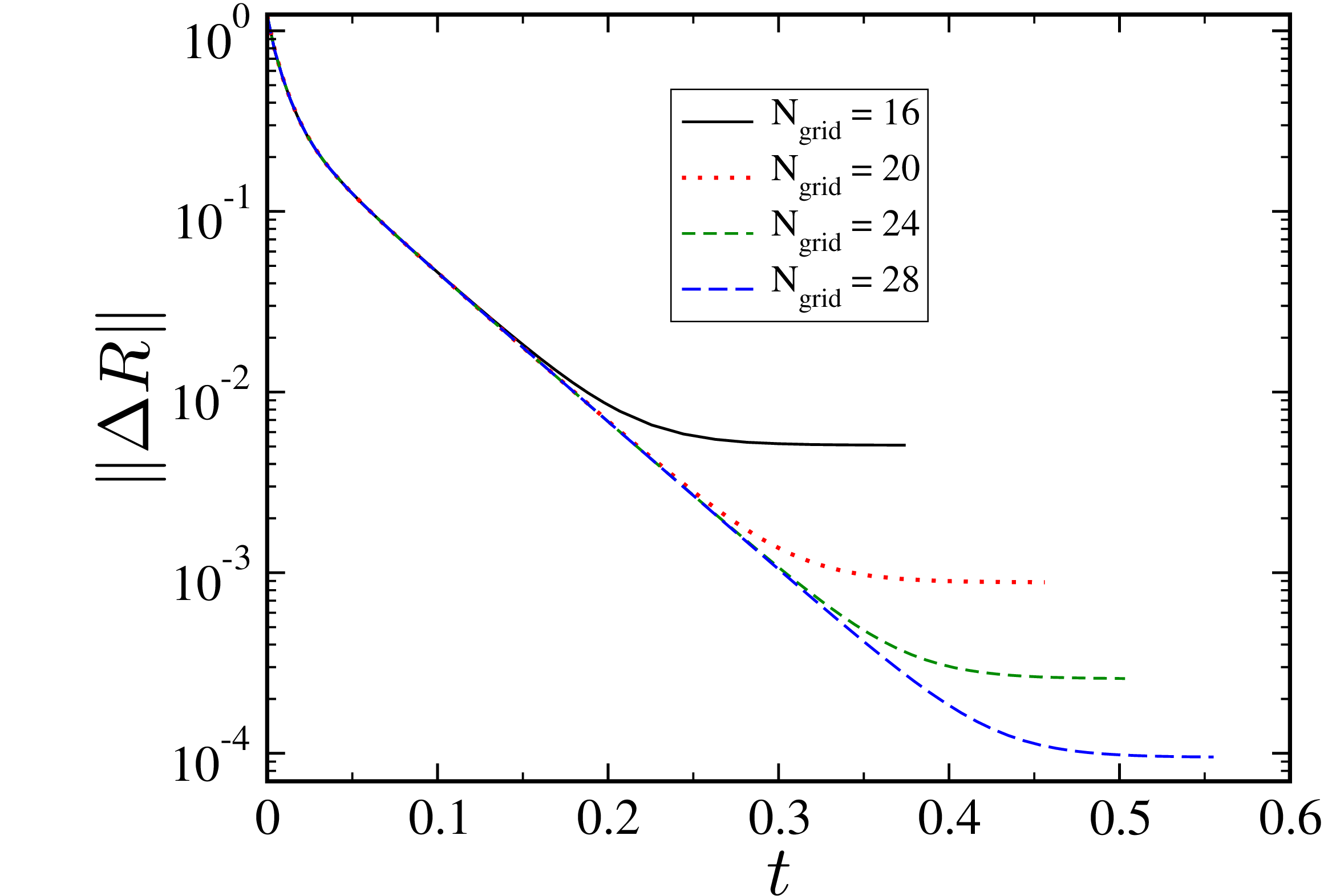}}
\caption{\label{f:SW-RFluctuation} This figure illustrates the norm
  $\|\,\Delta R\,\|$ defined in Eq.~(\ref{e:RFluctuationNorm}) that
  measures the average spatial variations of the Ricci scalar $R$ for
  the Ricci flow evolutions on the Seifert-Weber manifold.}
\end{figure}

\begin{figure}[!h]
  {\centering
    \includegraphics[height=0.32\textwidth]{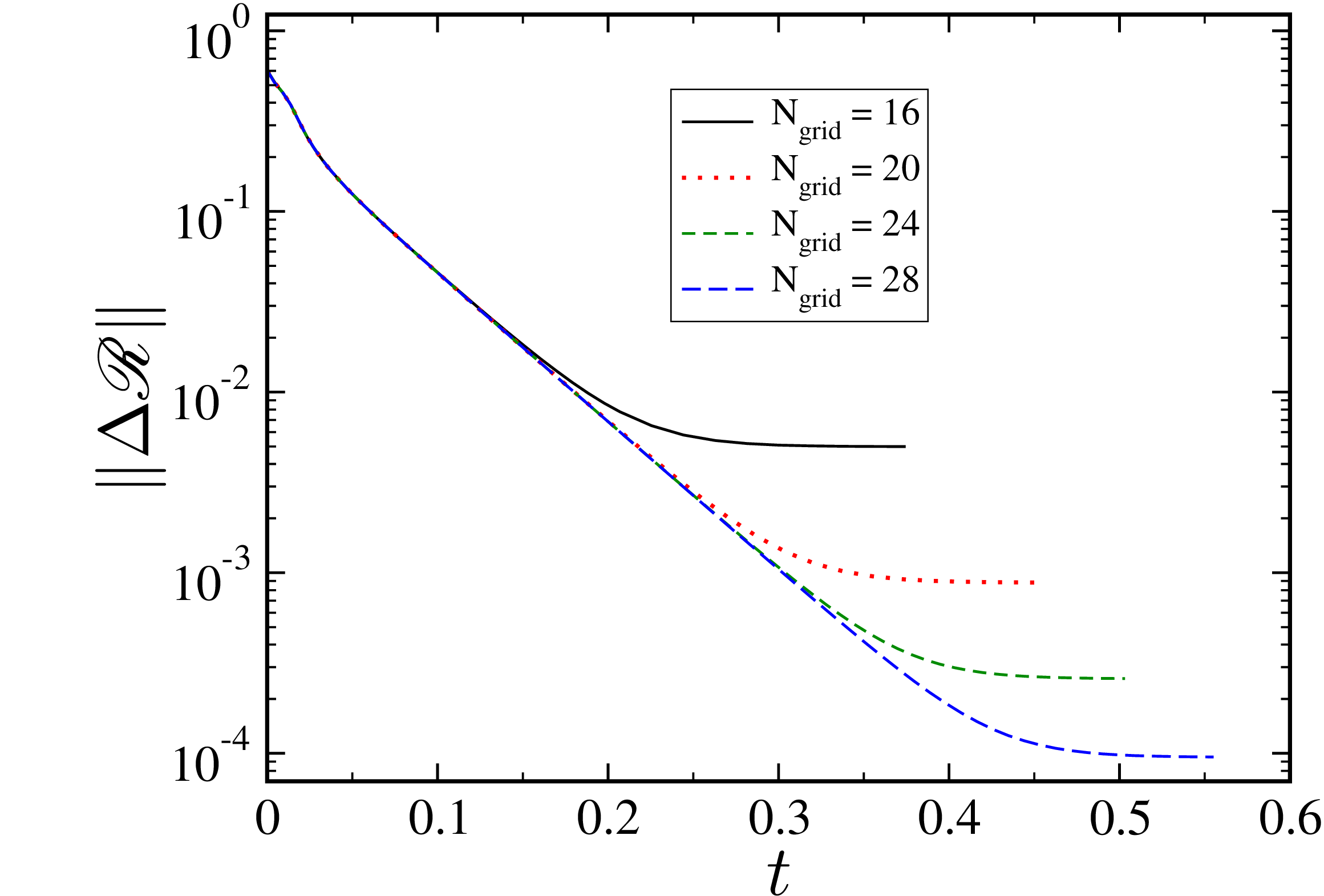}}
{  \caption{\label{f:SW-QFluctuation} This figure illustrates the norm
  $\|\,\Delta \mathscr{R}\,\|$ defined in
  Eq.~(\ref{e:CurvatureScaleFluctNorm0}) that measures the average
  spatial variations of the curvature scalar $\mathscr{R}$ for the
  Ricci flow evolutions on the Seifert-Weber manifold.  }}
\end{figure}

\begin{figure}[h]
  {\centering
    \includegraphics[height=0.32\textwidth]{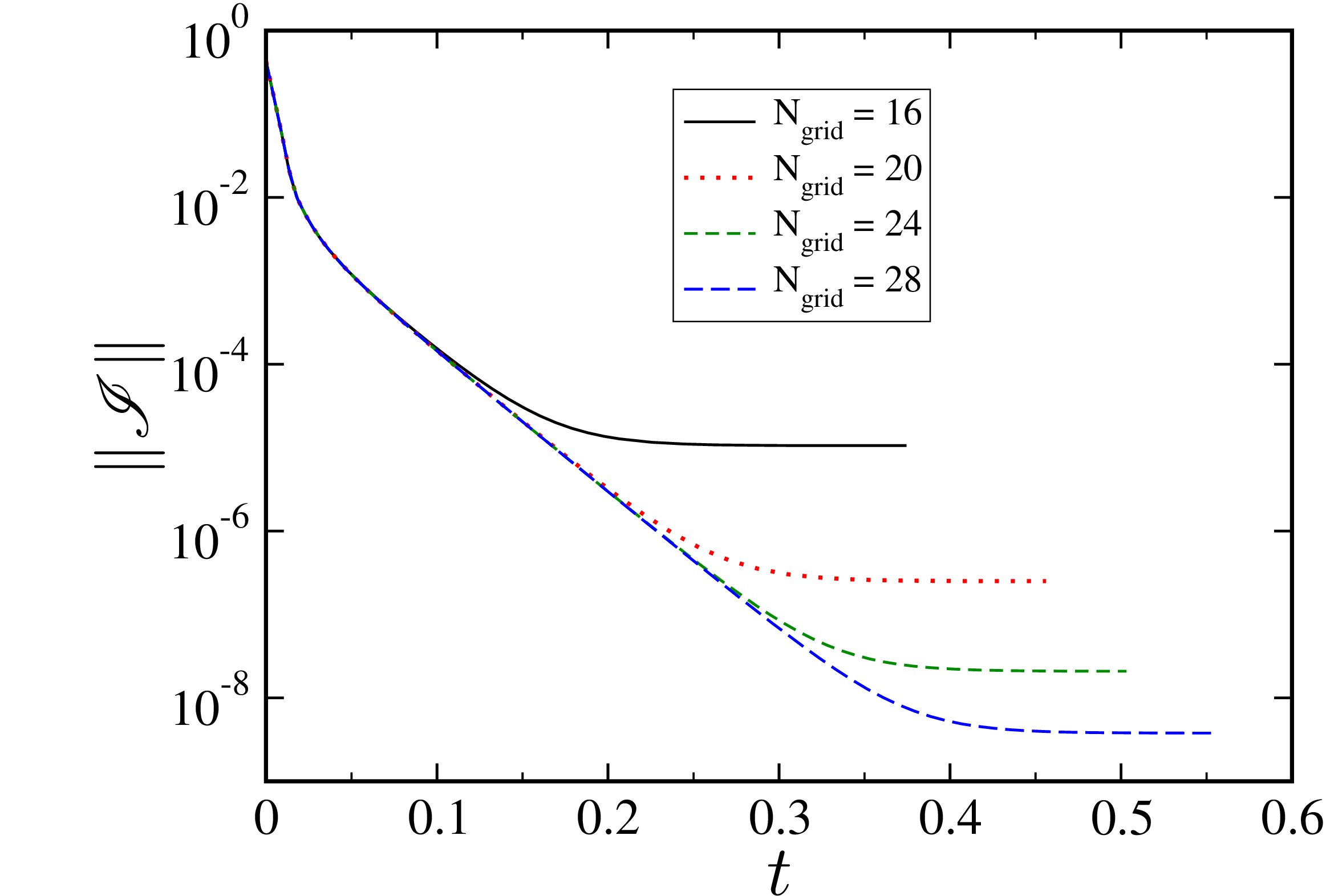}}
  { \caption{\label{f:SW-IsotropyNorm} This figure illustrates the norm
      $\|\,\mathscr{I}\,\|$ defined in Eq.~(\ref{e:IsotropyNorm}) that
      measures the fractional differences between the Ricci scalar
      norm $\llangle R\,\rrangle$ and the curvature scalar norm
      $\llangle \mathscr{R}\,\rrangle$ for the Ricci flow evolutions on
      the Seifert-Weber manifold. The equality $\mathscr{R}^{\,2}=R^2$ implies
      the isotropy of the spatial geometry.}}
\end{figure}

The Ricci flow equation on Seifert-Weber space has been solved for
this study using the reference metric $\tilde g_{ij}$ as initial
conditions.  These evolutions have been carried out using multi-cube
structures with a range of different $N_\mathrm{grid}$ grid points in
each spatial direction, $N_\mathrm{grid}=\{16,20,24,28\}$, so that the
numerical convergence of the results can be assessed.
Figure~\ref{f:SW-RFluctuation} illustrates the evolutions of the
spatial variations in the Ricci scalar, $\|\,\Delta R\,\|$, defined in
Eq.~(\ref{e:RFluctuationNorm}).  Figure~\ref{f:SW-QFluctuation}
illustrates the evolution of the spatial variations in the curvature
scalar, $\|\,\Delta\mathscr{R}\,\|$, defined in
Eq.~(\ref{e:CurvatureScaleFluctNorm0}).  And
Fig.~\ref{f:SW-IsotropyNorm} illustrates the evolution of the isotropy
norm, $\|\,\mathscr{I}\,\|$ defined in Eq.~(\ref{e:IsotropyNorm}).
These figures demonstrate that the Ricci flow evolutions on the
Seifert-Weber manifold are convergent and that they drive the metric
toward a homogeneous and isotropic state.

The Ricci flow equation is a parabolic system of equations for the
metric $g_{ij}$.  The numerical evolution of parabolic equations is
very inefficient because the maximum time step allowed for numerical
stability is extremely short, see Refs.~\cite{Canuto1988, Boyd2000}.
Consequently these Ricci flow evolutions take a lot of computational
resources.  For example, the evolutions on Seifert-Weber space took a
total of 10,267 cpu hours, or 513 cpu hours for each of the 20 cubic
regions used to represent this space. In order to improve the
efficiency of these computations a bootstrap method was developed for
the Ricci flow evolutions on the lens space $L(5,2)$, and the
Sixth-Turn space.  In this method the lowest resolution,
$N_\mathrm{grid}=16$, evolution is performed using the reference
metric $\tilde g_{ij}$ for that manifold as initial data.  This lowest
resolution evolution is continued until $\llangle \Delta R\,\rrangle$,
the spatial variations in the Ricci scalar curvature defined in
Eq.~(\ref{e:RFluctuation}), becomes constant (as seen for example in
Fig.~\ref{f:SW-RFluctuation} for the Seifert-Weber evolutions).  The
metric $g_{ij}$ that results from this initial evolution is then
interpolated onto the next higher resolution grid,
$N_\mathrm{grid}=20$.  These higher resolution data are used as
initial data for the next Ricci flow evolution which is continued
until $\llangle \Delta R\,\rrangle$ again becomes constant.  This
bootstrap process is repeated iteratively to produce the
$N_\mathrm{grid}=\{20,24,28\}$ resolution solutions to the Ricci flow
equations on the lens space $L(5,2)$ and on Sixth-Turn space.  Using
this bootstrap procedure reduces the total computational time by about
a factor of five: taking 101 cpu hours for each cubic region on the
lens space $L(5,2)$, and 107 cpu hours per cubic region on Sixth-Turn
space.

\begin{figure}[!h]
  {\centering
    \includegraphics[height=0.32\textwidth]{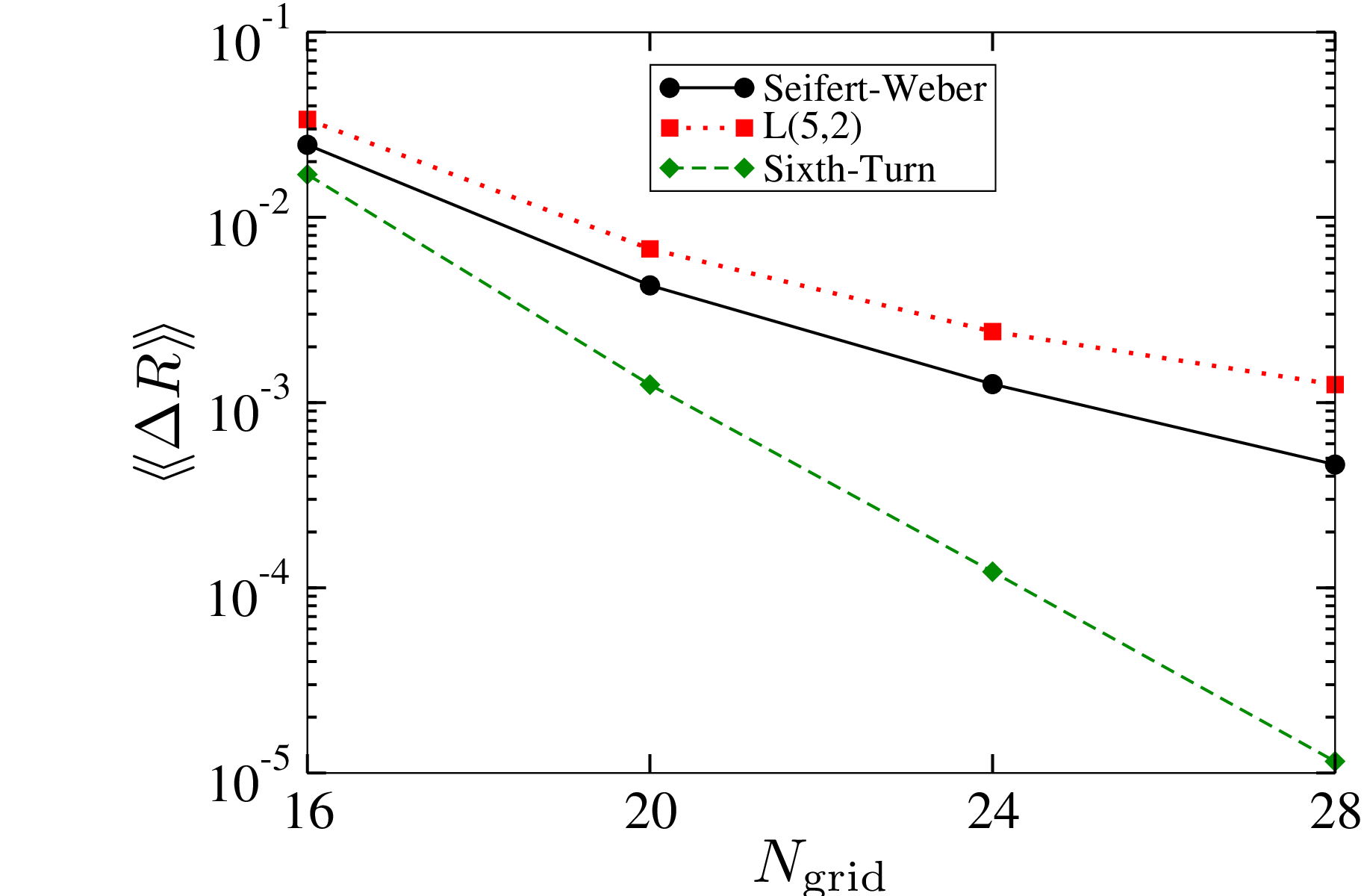}}
  { \caption{\label{f:DeltaRConvergence} This figure illustrates the
      convergence of $\llangle\,\Delta R\,\rrangle$ defined in
      Eq.~(\ref{e:RFluctuation}) as functions of the grid spacing
      parameter $N_\mathrm{grid}$ for the Ricci flow evolutions on the
      Seifert-Weber, L(5,2), and Sixth-Turn manifolds. }}
\end{figure}

\begin{figure}[!h]
  {\centering
    \includegraphics[height=0.32\textwidth]{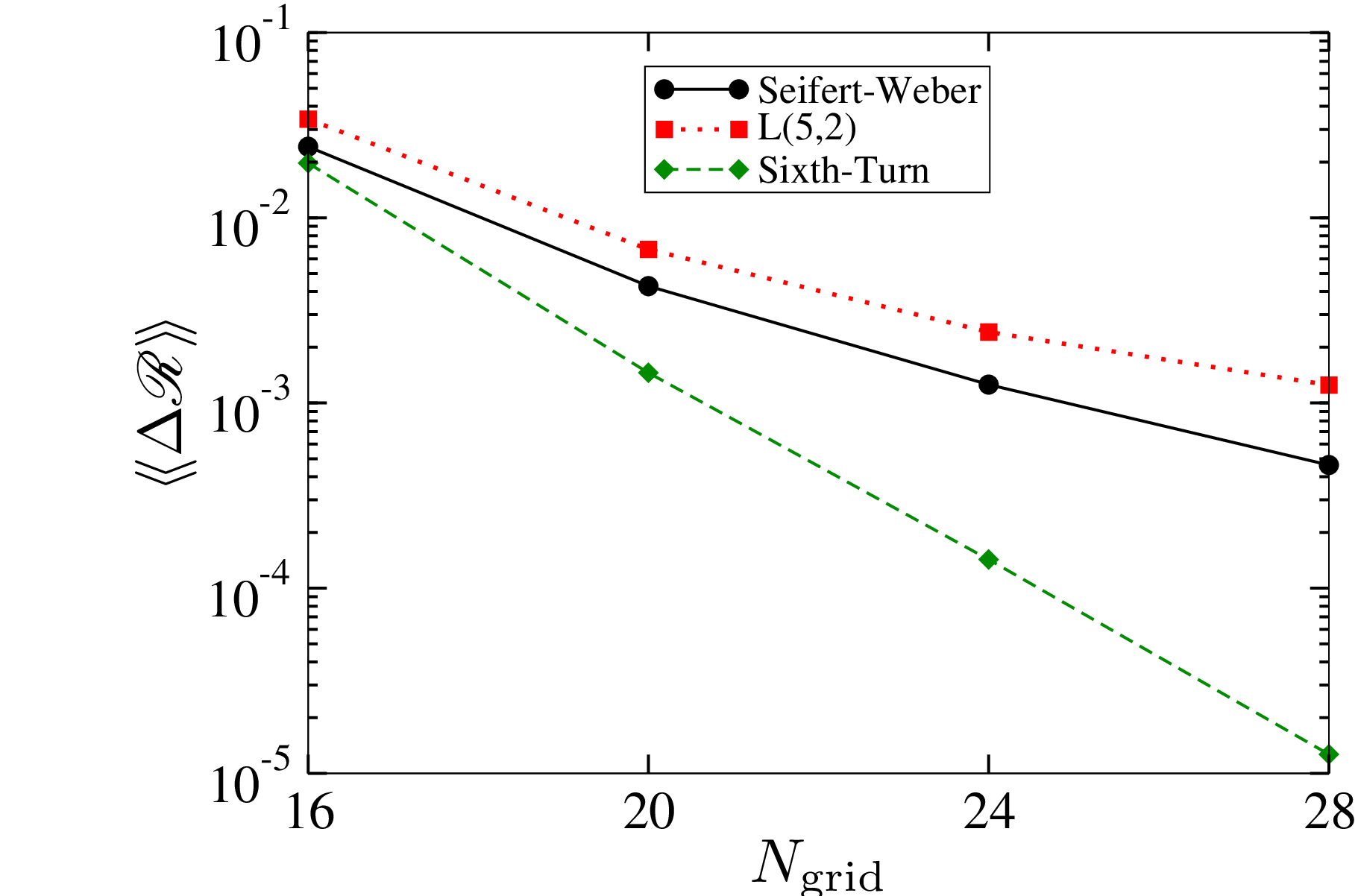}}
  { \caption{\label{f:DeltaQConvergence} This figure illustrates the
      convergence of $\llangle\,\Delta \mathscr{R}\,\rrangle$ defined in
      Eq.~(\ref{e:CurvatureScaleFluct}) as functions of the grid spacing
      parameter $N_\mathrm{grid}$ for the Ricci flow evolutions on the
      Seifert-Weber, L(5,2), and Sixth-Turn manifolds. }}
\end{figure}

\begin{figure}[!h]
  {\centering
    %    \includegraphics[height=0.32\textwidth]
    %{../Figures/IsotrophyConvergence.eps}}
    \includegraphics[height=0.32\textwidth] {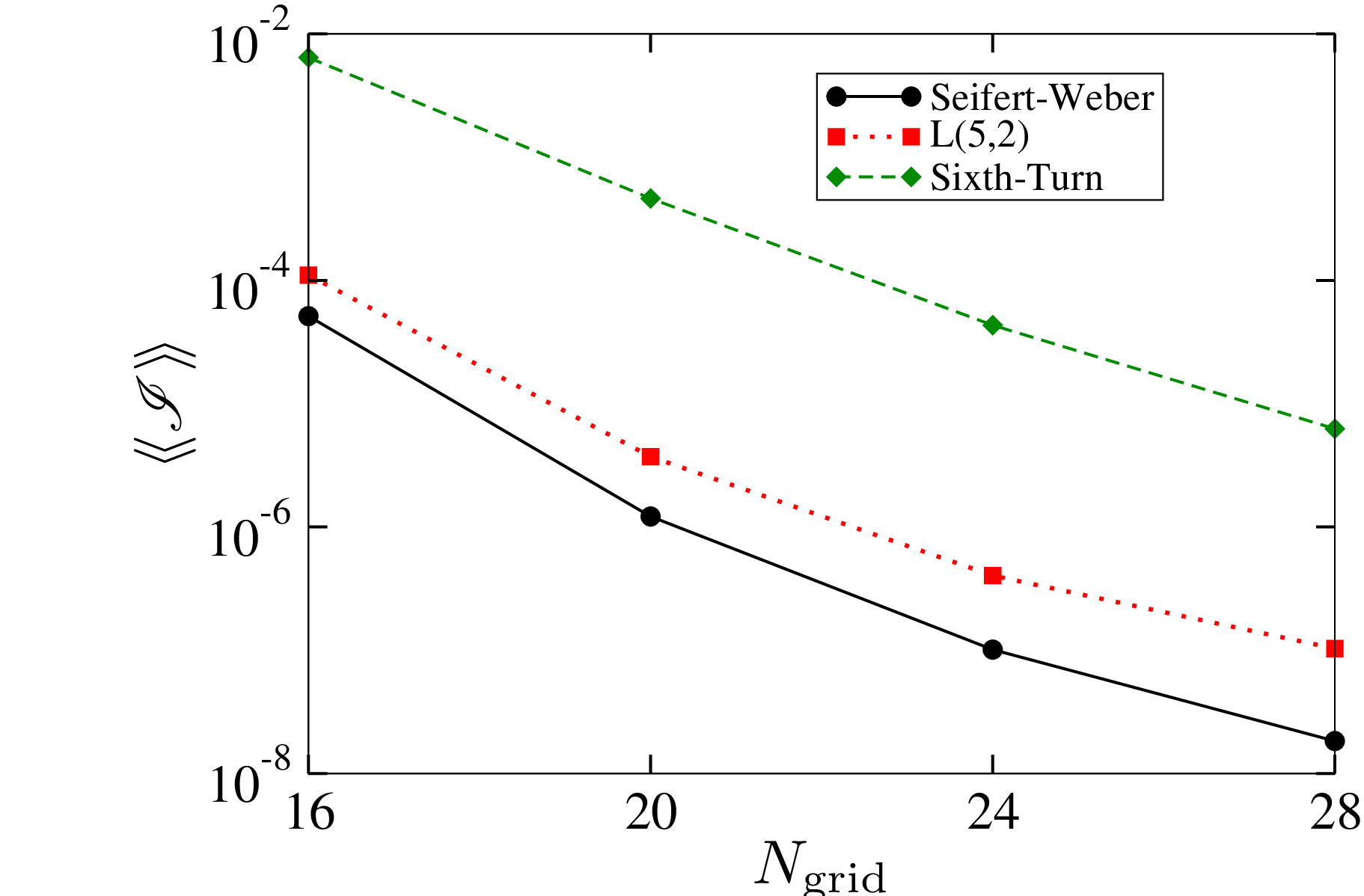}}
  { \caption{\label{f:IsotrophyConvergence} This figure illustrates
      the convergence of $\llangle\,\mathscr{I}\,\rrangle$
      defined in Eq.~(\ref{e:IsotropyMeasure}) as functions of the
      grid spacing parameter $N_\mathrm{grid}$ for the Ricci flow
      evolutions on the Seifert-Weber, L(5,2), and Sixth-Turn
      manifolds. }}
\end{figure}

Solving the Ricci flow equations on a sequence of grids with different
resolutions using this bootstrap method does not produce complete
evolution plots comparable to those in
Figs.~\ref{f:SW-RFluctuation}--\ref{f:SW-IsotropyNorm} for the
Seifert-Weber case.  The final Ricci flow metrics produced by this
process can, however, be determined and from them the numerical
convergence of the spatial variations in their curvature scalars,
$\llangle\,\Delta R\,\rrangle$ and $\llangle\,\Delta
\mathscr{R}\,\rrangle$, as well as the isotropy parameter
$\llangle\,\mathscr{I}\,\rrangle$ can be
evaluated. Figures~\ref{f:DeltaRConvergence}--\ref{f:IsotrophyConvergence}
compare these convergence results for the Seifert-Weber, the lens
space $L(5,2)$, and Sixth-Turn space.  These results show that the
Ricci flow evolutions are convergent toward geometries that are
spatially homogeneous and isotropic on these manifolds.
%

%%%%%%%%%%%%%%%%%%%%%%%%%%%%%%%%%%%%%%%%%%%%%%%%%%%%%%%%%%%%%%%%%%%%%%%%%%%%%%%
% Simple Initial Data
%%%%%%%%%%%%%%%%%%%%%%%%%%%%%%%%%%%%%%%%%%%%%%%%%%%%%%%%%%%%%%%%%%%%%%%%%%%%%%%
\section{Simple Initial Data}
\label{s:InitialData}
%%%%%%%%%%%%%%%%%%%%%%%%%%%%%%%%%%%%%%%%%%%%%%%%%%%%%%%%%%%%%%%%%%%%%%%%%%%%%%%

This section describes the method used in this study to find solutions
to the Einstein initial value equations.  The goal is to construct
initial data that can be used to evolve simple homogeneous
cosmological models on these manifolds.  The Einstein constraint
equations determine the allowed initial data for the spatial metric
$g_{ij}$ and extrinsic curvature $K_{ij}$ on a spacelike hypersurface.
The simplest solutions to these equations are determined by specifying
a conformal metric, $\tilde g_{ij}$, and a constant, $K$, that
functions as the trace of the extrinsic curvature.  The physical
metric $g_{ij}$ and the full extrinsic curvature $K_{ij}$ in this
approach are given by
\begin{eqnarray}
g_{ij} &=& \phi^4 \tilde g_{ij},
  \label{e:gmetricDef}\\
  K_{ij} &=& \tfrac{1}{3}\phi^4 \tilde g_{ij} K,
  \label{e:KDef}
\end{eqnarray}
where $\phi$ is a conformal factor.  This conformal factor is
determined by solving the constraint equations, which for this simple
case reduce to a second-order elliptic partial differential equation
for $\phi$.

The methods for solving the particular version of the Einstein
constraint equations used in this study were developed in
Ref.~\cite{Zhang2022} for manifolds with non-trivial topologies.  This
form of the initial value problem assumes that the average matter
content in these solutions is negligible compared to the cosmological
constant $\Lambda$.  With this assumption the Einstein constraint
equation reduces to
\begin{eqnarray}
  \tilde\nabla^i\tilde\nabla_i\phi &=& \textstyle\frac{1}{8}\phi\,
  \tilde R +\textstyle\frac{1}{12}\phi^5\, (K^2-3\Lambda),
  \label{e:CMCConstraintS} 
\end{eqnarray}
where $\tilde \nabla_i$ and $\tilde R$ are the covariant derivative
and Ricci scalar curvature respectively associated with the conformal
metric $\tilde g_{ij}$.  The conformal metrics $\tilde g_{ij}$ used in
this study are the solutions to the Ricci flow equation described in
Sec.~\ref{s:RicciFlowNumericalResults}.

The integral of the left side of Eq.~(\ref{e:CMCConstraintS}) vanishes
on any compact manifold without boundary. Therefore the constants $K$
and $\Lambda$ must be chosen in a way that makes it possible for the
integral of the right side of this equation to vanish as well.
Convenient choices for these constants produce solutions to
Eq.~(\ref{e:CMCConstraintS}) with $\phi\approx 1$.  An acceptable
choice can be identified by setting $\phi=1$ in the expression on the
right side of Eq.~(\ref{e:CMCConstraintS}) and integrating over the
manifold.  Setting this integral to zero results in the following
constraint on the values of $K$ and $\Lambda$,
\begin{equation}
  K^2 - 3\Lambda = - \textstyle\frac{3}{2}\llangle \tilde R\, \rrangle,
  \label{e:CMCParameterChoices}
\end{equation}
where $\llangle \tilde R\, \rrangle$ is the average value of the
conformal scalar curvature $\tilde R$,
\begin{equation}
  \llangle \tilde R\, \rrangle
  = \frac{\int \sqrt{\tilde g}\,\tilde R\, d^{\,3}x}
  {\int \sqrt{\tilde g}\, d^{\,3}x},
  \label{e:AvarageRtildeDef}
\end{equation}
where $\tilde g$ represents the determinant of the conformal metric
$\tilde g_{ij}$.  This choice transforms Eq.~(\ref{e:CMCConstraintS})
into the form 
\begin{eqnarray}
  \tilde\nabla^i\tilde\nabla_i\phi &=& \textstyle\frac{1}{8}\phi\,
  \left(\tilde R - \phi^4 \llangle \tilde R\,\rrangle\right).
  \label{e:CMCConstraintSAlt}
\end{eqnarray}
This equation has the exact solution $\phi=1$ in the constant scalar
curvature case $\tilde R =\llangle \tilde R\,\rrangle$.

The Ricci scalar curvature $R$ determined from the physical metric
$g_{ij}$ is related to the Ricci scalar curvature $\tilde R$
associated with the conformal metric $\tilde g_{ij}$ by the expression
\begin{eqnarray}
R &=& \phi^{-4}\, \tilde R - 8 \phi^{-5}\, \tilde \nabla^i\tilde\nabla_i\phi.
  \label{e:ConformalRicciScalar}
\end{eqnarray}
The Hamiltonian constraint, $\mathcal{H}$, for the initial value
problem considered here is given by the expression
\begin{equation}
  \mathcal{H}=R +\textstyle\frac{2}{3}\left(K^2 - 3\Lambda\right) = R
  - \llangle\tilde R\rrangle,
\end{equation}
(see Ref.~\cite{Zhang2022}).  Solutions to the Einstein constraint
Eq.~(\ref{e:CMCConstraintSAlt}) therefore have physical Ricci scalar
curvatures given by
\begin{eqnarray}
  R&=&\llangle\tilde R\,\rrangle.
  \label{e:ConstantRCondition}
\end{eqnarray}
The accuracy of numerically determined
solutions to the initial value problem can be evaluated 
using the Hamiltonian constraint norm $\|\,\mathcal{H}\,\|$ defined
by,
\begin{equation}
  \|\,\mathcal{H}\,\|^2 = \mathcal{V}^{-1}\int \left(R -
  \llangle\tilde R\, \rrangle\right)^2 \!\sqrt{g}\, d^{\,3}x.
  \label{e:EinsteinConstraint}
\end{equation}

The conformal metrics $\tilde g_{ij}$ used to solve the initial value
problem for this study are the metrics constructed in
Sec.~\ref{s:RicciFlowNumericalResults} by solving the Ricci flow
equation.  These Ricci flow metrics have the property that the spatial
variations in the Ricci scalar, $\llangle\,\Delta R\, \rrangle$, are
very small, see Fig.~\ref{f:DeltaRConvergence}.  The Hamiltonian
constraint norm $\|\mathcal{H}\,\|$ is identical to the spatial
variations of the Ricci scalar curvature defined in
Eq.~(\ref{e:RFluctuation}): $\|\,\mathcal{H}\,\|=\llangle\,\Delta
R\,\rrangle$. Therefore the Ricci flow metrics are themselves
solutions to the initial value Eq.~(\ref{e:CMCConstraintS}) with
$\phi=1$.  Table~\ref{t:TableI} lists the average values of the
physical Ricci scalar curvature, $\llangle R\,\rrangle$, the physical
volume $\mathcal{V}$, and the cosmological constant $\Lambda$ for the
initial spatial geometries constructed for each of the manifolds
included in this study.  The volumes $\mathcal{V}$ measure the
physical ``sizes'' of the manifolds in the length-scale units of our
code, and can therefore be used to calibrate the sizes of the
curvatures of the geometries.  The graphs in
Figs.~\ref{f:DeltaRConvergence} and \ref{f:DeltaQConvergence} show
that these Ricci flow geometries are homogeneous, and the graph in
Fig.~\ref{f:IsotrophyConvergence} shows that they are also isotropic.
\begin{table}[!hbt]
  \caption{Physical properties of the initial data for the manifolds
    included in this study: the average scalar curvature $\llangle
    R\,\rrangle$ defined in Eq.~(\ref{e:RAverage}), the physical
    volume $\mathcal{V}$ defined in Eq.~(\ref{e:Volume}), and the
    cosmological constant $\Lambda$ determined by
    Eq.~(\ref{e:CMCParameterChoices}) for the geometries with $K=-1$
    constructed on each manifold.
    \label{t:TableI} }
  \begin{center}
  \begin{tabular}{lrlrr}
    Manifold && $\!\!\!\!\!\!\!\!\!\llangle R\, \rrangle$
    & $\qquad\mathcal{V}\qquad$  & $\qquad\Lambda\qquad$
 \\
 \hline
 \vspace{-7pt}\\
 Seifert-Weber   &   \qquad -4.&\!\!\!8467 & \, 15.4254  &  -2.0900 \\
 $L(5,2)$        &   \qquad 3.&\!\!\!9873 & \, 7.2874   &   2.3270 \\
 Sixth-Turn      &   \qquad -1.&\!\!\!7538$\times 10^{-8}$ & \, 6.2804 & 0.3333 \\
 \hline
  \end{tabular}
  \end{center} 
\end{table}
%

%%%%%%%%%%%%%%%%%%%%%%%%%%%%%%%%%%%%%%%%%%%%%%%%%%%%%%%%%%%%%%%%%%%%%%%%%%%%%%%
\section{Simple Cosmological Models}
\label{s:CosmologicalModels}
%%%%%%%%%%%%%%%%%%%%%%%%%%%%%%%%%%%%%%%%%%%%%%%%%%%%%%%%%%%%%%%%%%%%%%%%%%%%%%%
This section describes the methods used in this study to construct
simple homogeneous cosmological models by solving the Einstein
evolution equations using the initial data prepared in
Secs.~\ref{s:RicciFlowNumericalResults} and \ref{s:InitialData}. The
spatial metrics, $g^0_{ij}$, produced by those solutions to the
Einstein constraint equations are used in two ways.  First, they are
used to construct the initial data for the spacetime metric
$\psi^0_{ab}$:
\begin{equation}
  ds^2=\psi^0_{ab}\,dx^adx^b = - dt^2 + g^0_{ij}\,dx^idx^j,
  \label{e:psi0Def}
\end{equation}
where the indices from the first part of the alphabet, e.g. $a$, $b$, $c$,
$...$, range over the four-dimensional spacetime coordinates, while the
indices from the later parts of the alphabet, e.g. $i$, $j$, $k$, $...$, range
over the three-dimensional spatial coordinates on each t = constant
hypersurface.  Second, the metric $\psi^0_{ab}$ is also used to define
the time-independent spacetime reference metric,
$\tilde\psi_{ab}\equiv\psi^0_{ab}$ that defines the differential
structure in these four-dimensional manifolds.
  
The representation of the Einstein equation used to perform the
numerical evolutions in this study is the covariant first-order
symmetric-hyperbolic representation developed in
Ref.~\cite{Lindblom2014}.  The dynamical fields in this representation
are the spacetime metric, $\psi_{ab}$, its time derivative,
$\Pi_{ab}=-t^c\tilde\nabla_c\psi_{ab}$, and its spatial derivatives
$\Phi_{iab}=\tilde\nabla_i\psi_{ab}$, where $t^c$ is the future
directed timelike unit normal ($t^a t^b \psi_{ab}=-1$) to the $t =$
constant hypersurfaces.  The derivative, $\tilde\nabla_a$, used in
these definitions is the torsion-free covariant derivative compatible
with the reference metric: $\tilde\nabla_c\tilde\psi_{ab}=0$. The
basic structure of this representation of Einstein's equation is
summarized more completely in
Appendix~\ref{s:SymmetricHyperbolicSystem}. The values of the
dynamical fields on the initial spacelike hypersurface used to start
the evolutions in this study are given by,
\begin{eqnarray}
  \psi^0_{ab}&=& \tilde\psi_{ab},
  \label{e:Initialpsi}\\
  \Pi^0_{tt}&=&-2K^0,
  \label{e:InitialPitt}\\
  \Pi^0_{tk}&=&0,
  \label{e:InitialPitk}\\
  \Pi^0_{ij}&=&\tfrac{2}{3}K^0 g^0_{ij},
  \label{e:InitialPiij}\\
  \Phi^0_{iab}&=&0,
  \label{e:InitialPhi}
\end{eqnarray}
where $K^0$ is the trace of the extrinsic curvature used to compute
the initial data in Sec.~\ref{s:InitialData}. 

\begin{figure}[!h]
  {\centering
      \includegraphics[height=0.32\textwidth]{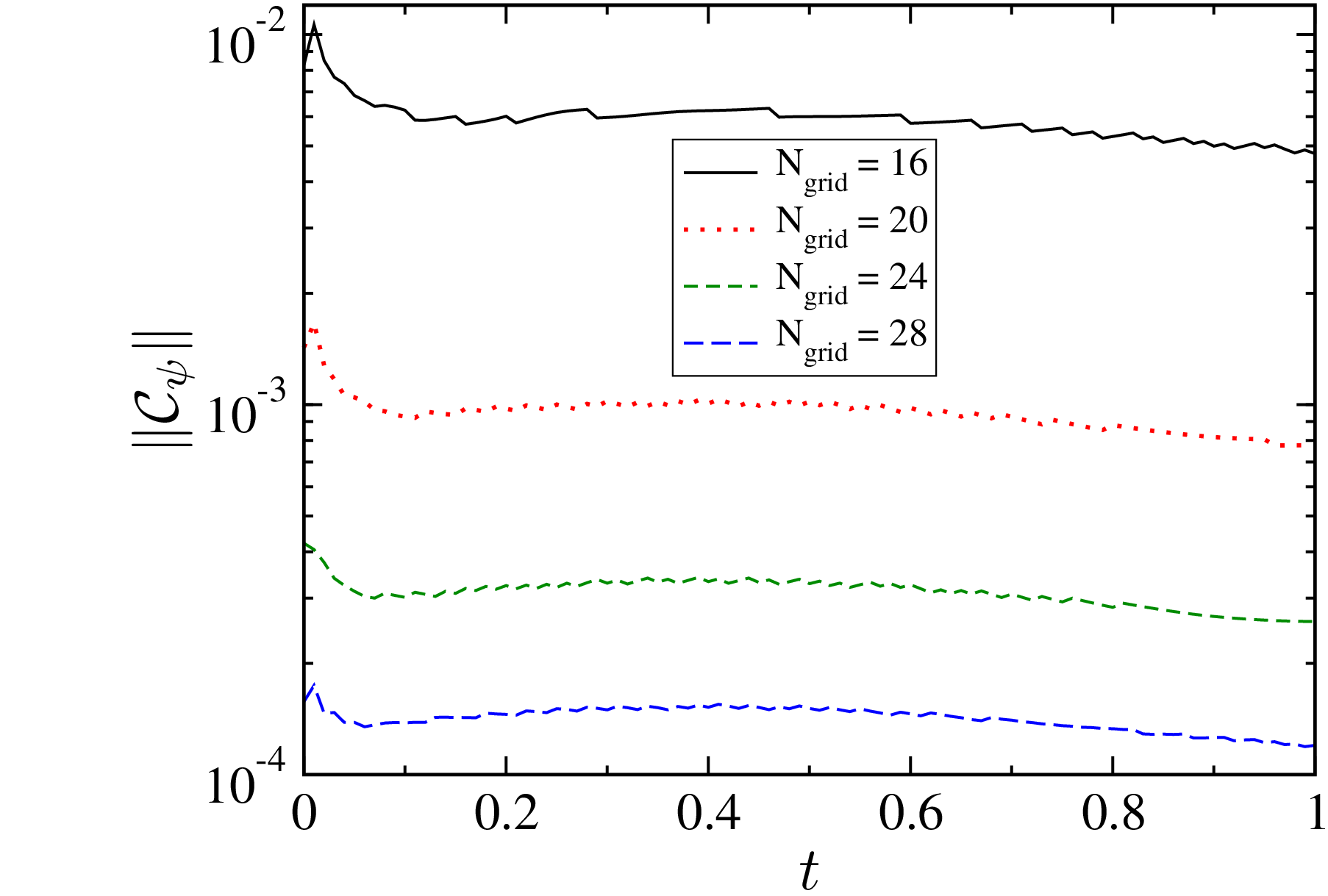}}
  { \caption{\label{f:SW-Constraints} This figure illustrates the
      dimensionless constraint norm $\|\,\mathcal{C}_\psi\,\|$ defined
      in Eq.~(\ref{e:CpsiNormDef}) that measures the constraint
      violations for the solutions to the Einstein evolution equations
      on the Seifert-Weber manifold.  }}
\end{figure}

\begin{figure}[!h]
  {\centering
    \includegraphics[height=0.32\textwidth]{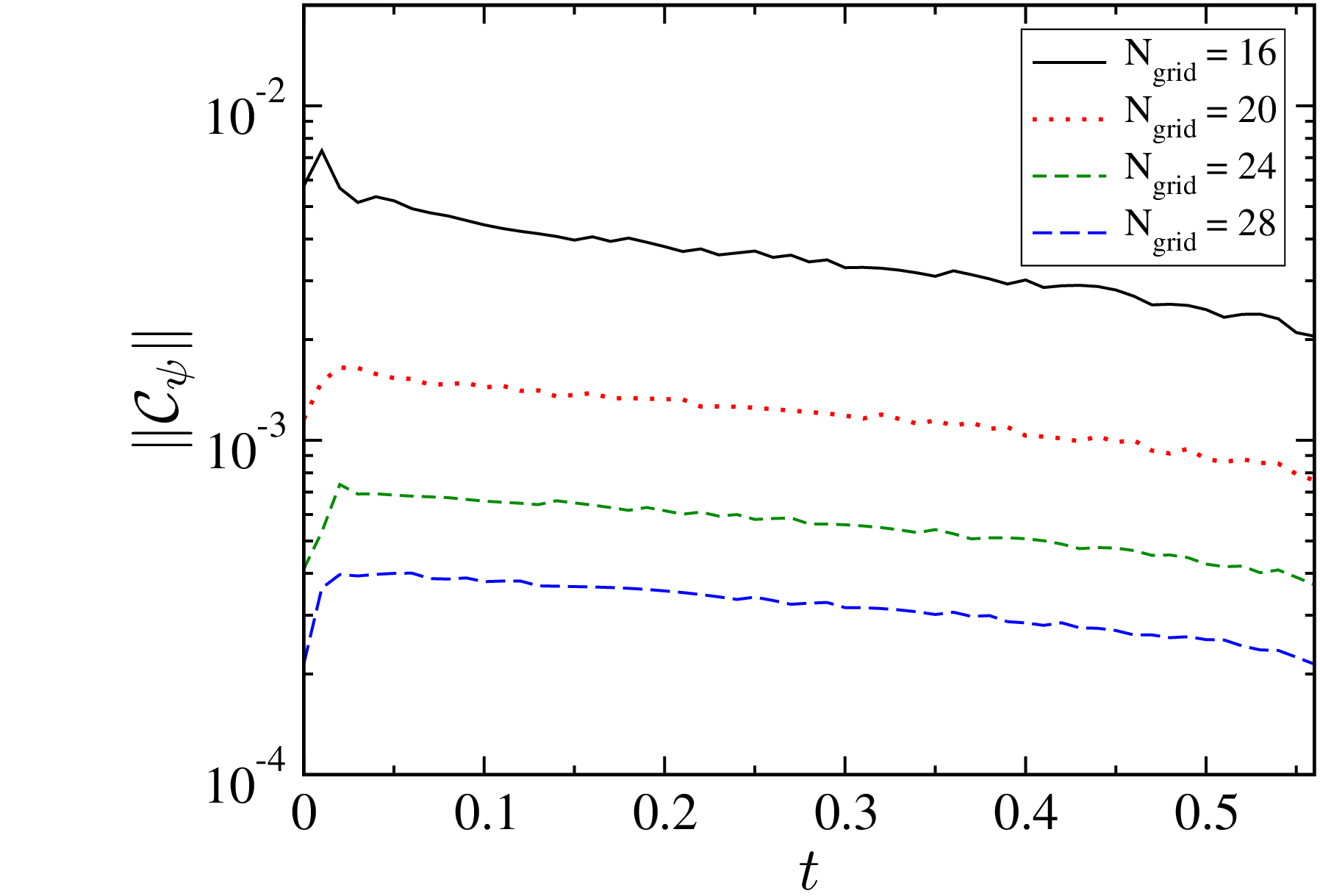}}
  { \caption{\label{f:L52-Constraints} This figure illustrates the
      dimensionless constraint norm $\|\,\mathcal{C}_\psi\,\|$ defined
      in Eq.~(\ref{e:CpsiNormDef}) that measures the constraint
      violations for the solutions to the Einstein evolution equations
      on the $L(5,2)$ manifold.  }}
\end{figure}

\begin{figure}[!h]
  {\centering
    \includegraphics[height=0.32\textwidth]{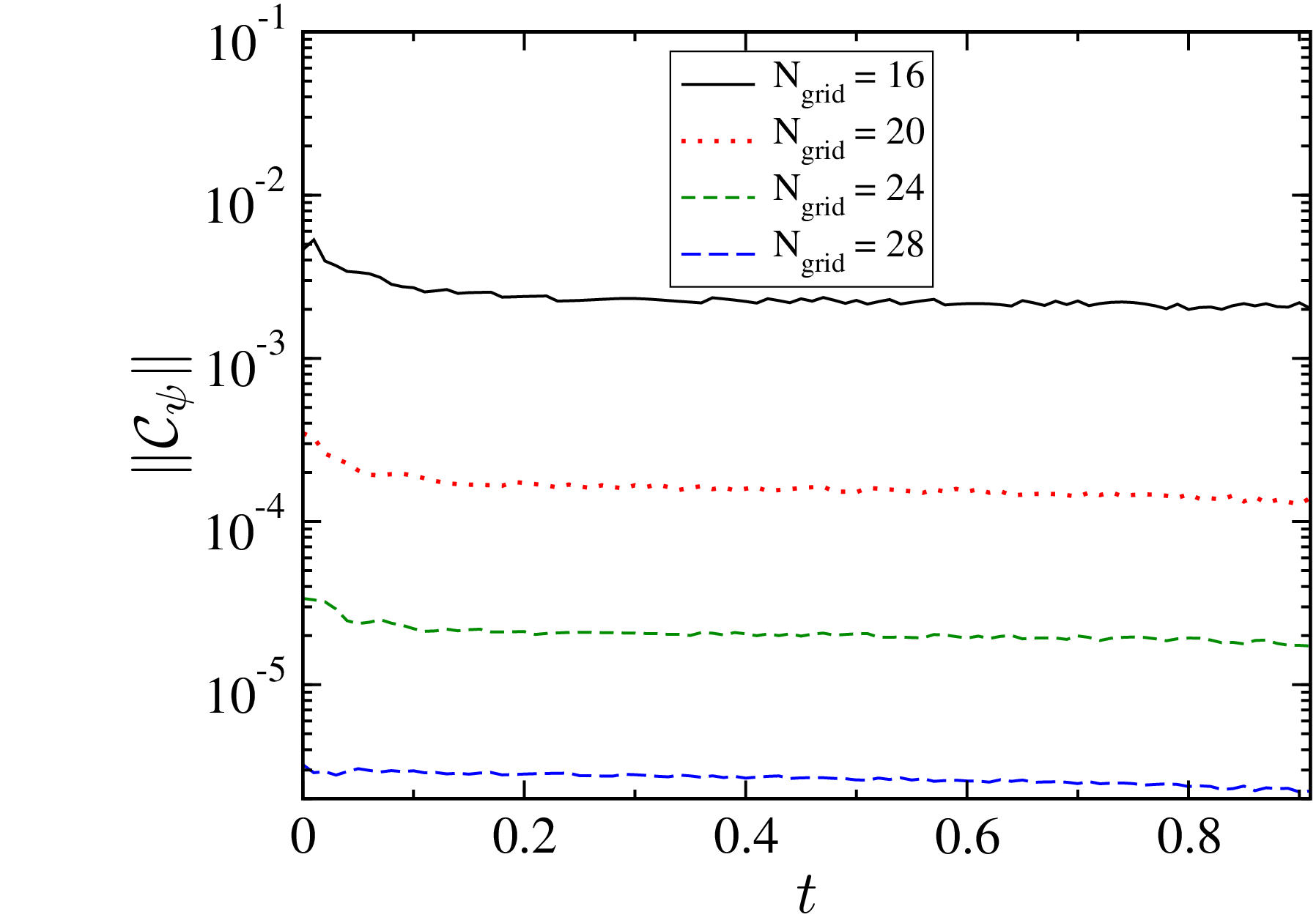}}
  { \caption{\label{f:E5-Constraints} This figure illustrates the
      dimensionless constraint norm $\|\,\mathcal{C}_\psi\,\|$ defined
      in Eq.~(\ref{e:CpsiNormDef}) that measures the constraint
      violations for the solutions to the Einstein evolution equations
      on the Sixth-Turn space.  }}
\end{figure}

The first-order symmetric-hyperbolic representation of the Einstein
system has a large number of constraints, which are described in some
detail in Appendix~\ref{s:SymmetricHyperbolicSystem}.  The overall
magnitude of these constraints is summarized by a quantity
$\mathcal{C}_\psi$ defined in Eq.~(\ref{e:CpsiDef}).  The
dimensionless norm $\|\,\mathcal{C}_\psi\,\|$ defined in
Eq.~(\ref{e:CpsiNormDef}) provides a useful measure of the fractional
constraint violation errors in these cosmological models.
Figure~\ref{f:SW-Constraints} illustrates
$\|\,\mathcal{C}_\psi(t)\,\|$ for the numerical evolutions at each of
the numerical resolutions $N_\mathrm{grid}=\{16,20,24,28\}$ on
Seifert-Weber space.  This figure shows that these numerical
evolutions are convergent solutions to Einstein's
equation. Figures~\ref{f:L52-Constraints} and \ref{f:E5-Constraints}
give the analogous results for the lens space $L(5,2)$ and for
Sixth-Turn space, demonstrating that these evolutions have produced
numerically convergent cosmological models as well.

\begin{figure}[!h]
  {\centering
    \includegraphics[height=0.32\textwidth]{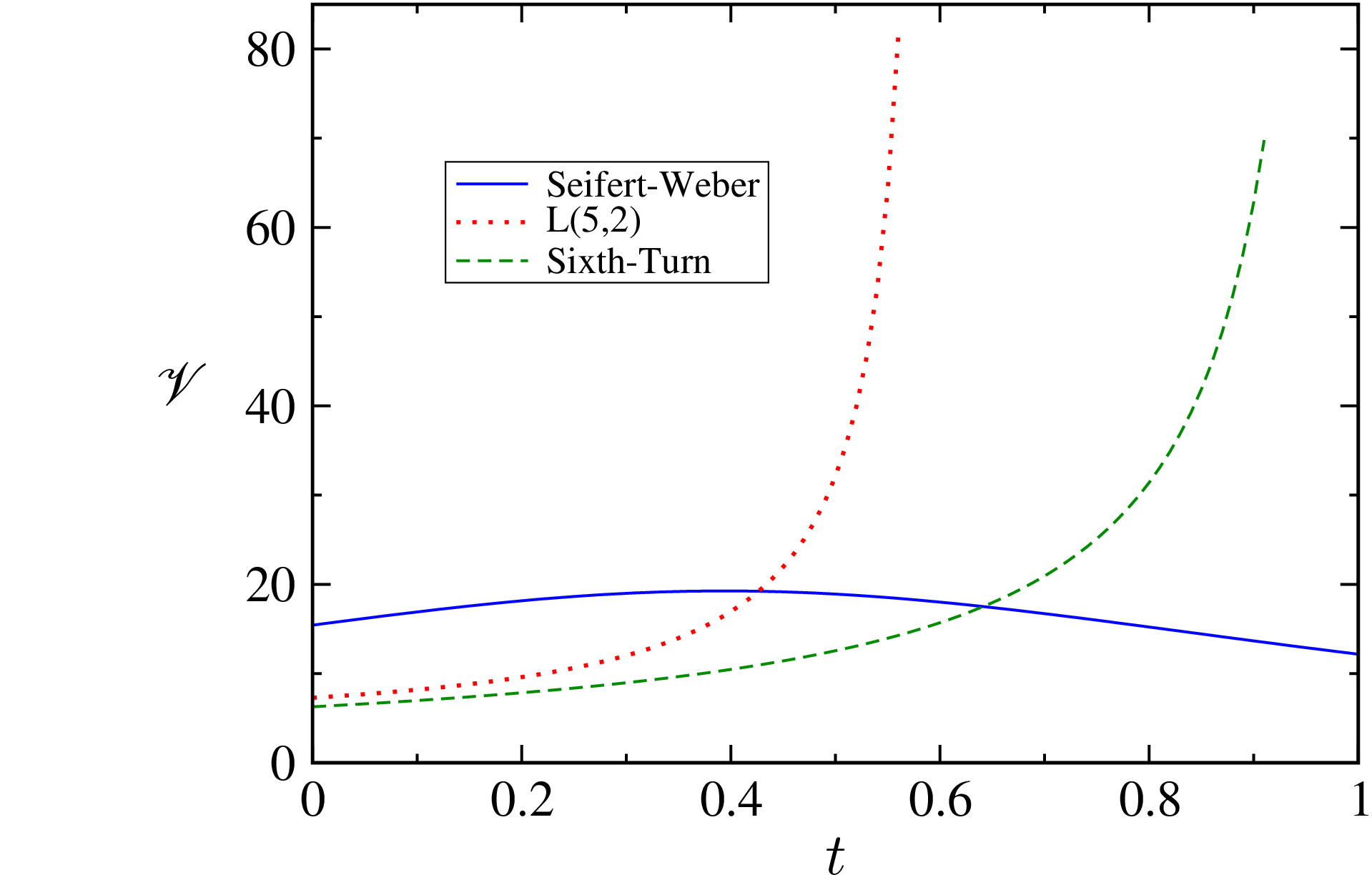}}
  { \caption{\label{f:Volume} This figure illustrates the evolution of
      the volume $\mathscr{V}(t)$ defined in Eq.~(\ref{e:Volume}) for
      the $N_\mathrm{grid} = 28$ solutions to Einstein's evolution
      equation on the Seifert-Weber, $L(5,2)$ and the Sixth-Turn
      manifolds.}}
\end{figure}

The properties of the cosmological models resulting from the numerical
evolutions performed for this study are most easily analyzed by
expressing the resulting spacetime metric, $\psi_{ab}$, in the
three-plus-one ADM form (see Ref.~\cite{ADM1962}):
\begin{eqnarray}
  ds^2 &=& \psi_{ab}\,dx^adx^b\nonumber\\
  &=& -N^2dt^2 + g_{ij}(dx^i+N^idt)(dx^j+N^jdt),\qquad
  \label{e:ADMMetric}
\end{eqnarray}
where $N$ is the lapse, $N^i$ the shift, and $g_{ij}$ the spatial
metric on each $t=$ constant hypersurface.  The initial data for these
cosmological models are evolved numerically until the volumes of the
$t=$ constant hypersurfaces increase by about a factor of ten in the
lens space $L(5,2)$ and in the Sixth-Turn space cases.  The volume in
Seifert-Weber space increases initially in these numerical evolutions,
but then contracts after reaching a maximum before expanding by a
factor of ten.  The evolution in this Seifert-Weber case was continued
for approximately the same length of coordinate time as the other two
cases.  Figure~\ref{f:Volume} illustrates $\mathcal{V}(t)$ for the
highest resolution evolutions, $N_\mathrm{grid}=28$, on these
manifolds.

\begin{figure}[!h]
  {\centering
    \includegraphics[height=0.32\textwidth]{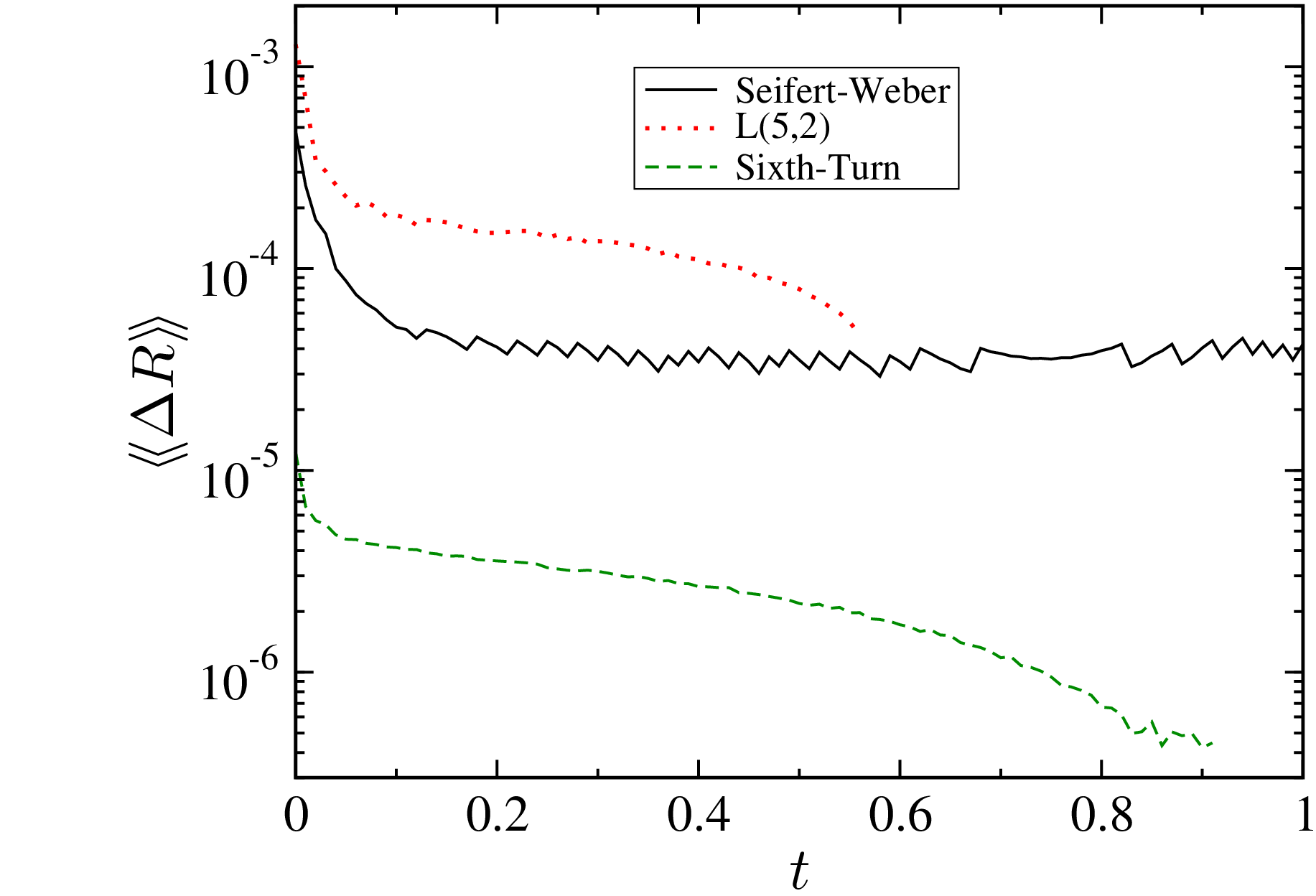}}
  { \caption{\label{f:RFluctuationEv} This figure illustrates the
      evolution of the norm $\llangle\,\Delta R(t)\,\rrangle$ defined in
      Eq.~(\ref{e:RFluctuation}) that measures the spatial
      variations in the Ricci scalar curvature $R$ for the
      $N_\mathrm{grid}=28$ solutions to the Einstein evolution equations on the
      Seifert-Weber, $L(5,2)$, and the Sixth-Turn manifolds.  }}
\end{figure}

\begin{figure}[!h]
  {\centering
    \includegraphics[height=0.32\textwidth]{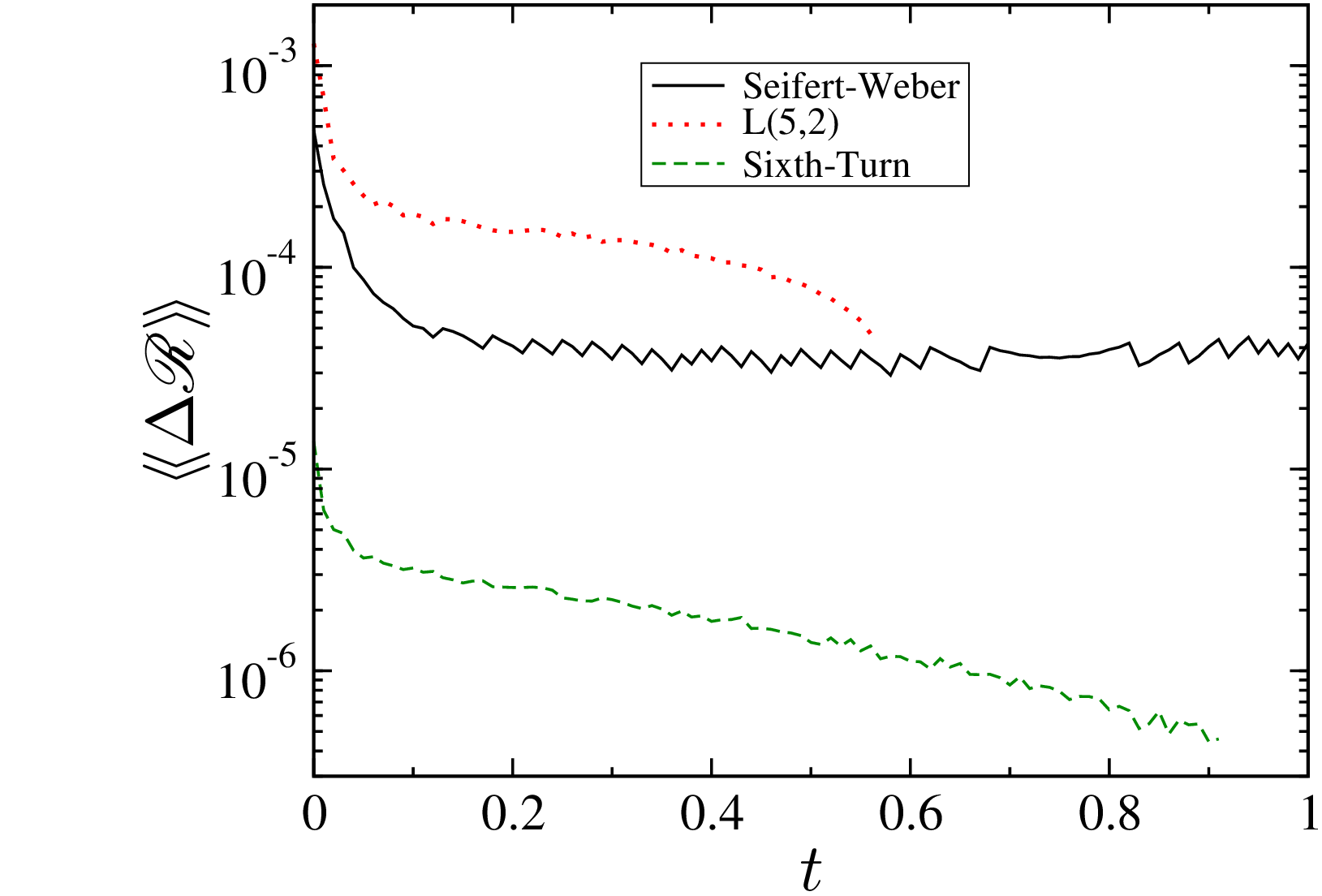}}
  { \caption{\label{f:QFluctuationEv} This figure illustrates the
      evolution of the norm $\llangle\,\Delta
      \mathscr{R}(t)\,\rrangle$ defined in
      Eq.~(\ref{e:CurvatureScaleFluct}) that measures the spatial
      variations in the curvature scalar $\mathscr{R}$ for the
      $N_\mathrm{grid}=28$ solutions to the Einstein evolution
      equations on the Seifert-Weber, $L(5,2)$, and the Sixth-Turn
      manifolds.  }}
\end{figure}

\begin{figure}[!h]
  {\centering
    \includegraphics[height=0.32\textwidth]{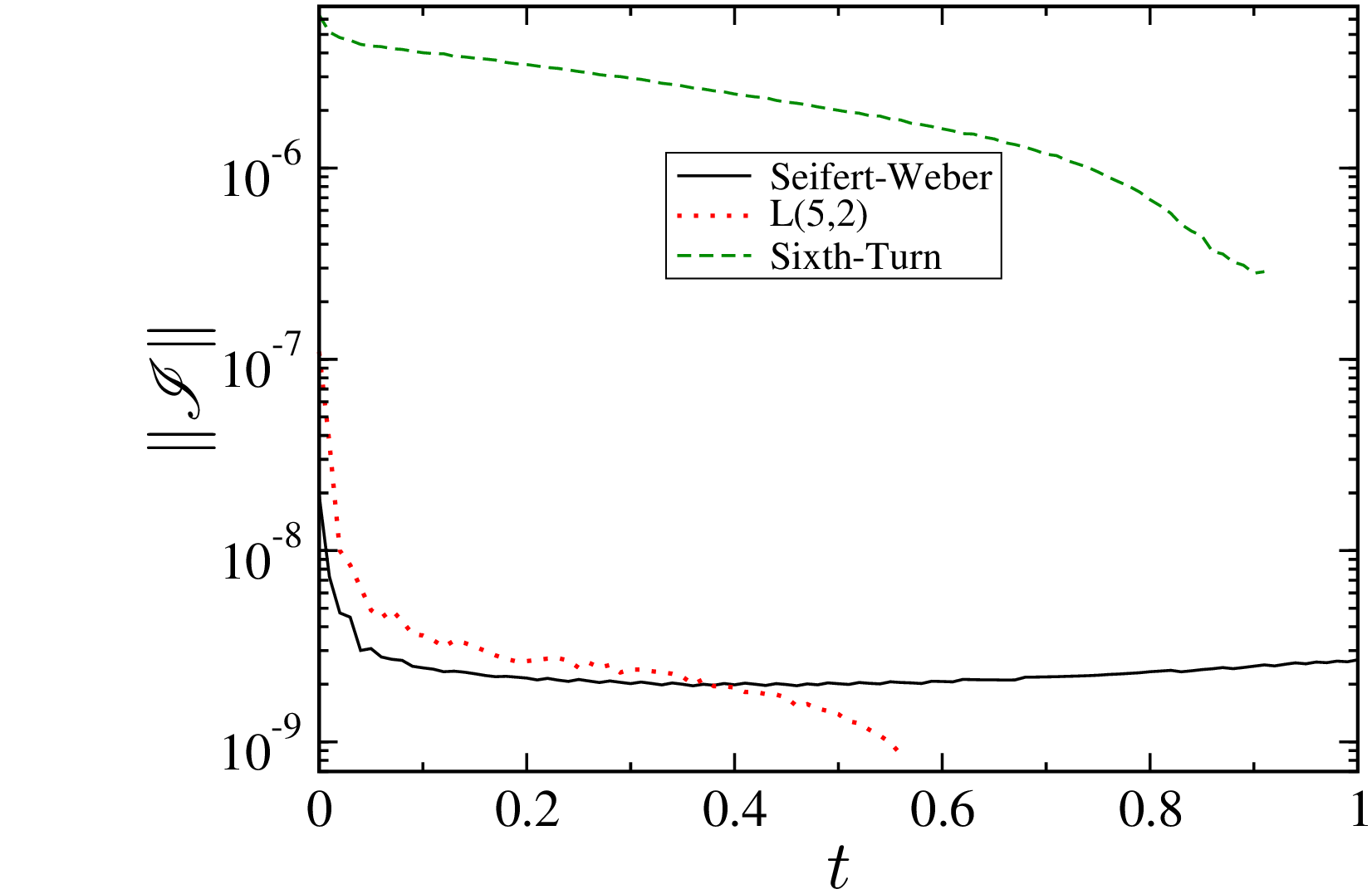}}
  { \caption{\label{f:IsotrophyEv} This figure illustrates the
      evolution of the norm $\llangle\,\mathscr{I}(t)\,\rrangle$
      defined in Eq.~(\ref{e:IsotropyMeasure}) that measures the
      isotropy of the spatial geometry for the $N_\mathrm{grid}=28$
      solutions to the Einstein evolution equations on the
      Seifert-Weber, $L(5,2)$, and the Sixth-Turn manifolds.  }}
\end{figure}

The homogeneity and isotropy of the geometries of the $t=$ constant
hypersurfaces in these simple cosmological models have been evaluated
by monitoring the norms $\llangle\,\Delta R(t)\,\rrangle$,
$\llangle\,\Delta \mathscr{R}(t)\,\rrangle$, and
$\llangle\,\mathscr{I}(t)\,\rrangle$ defined in
Eqs.~(\ref{e:RFluctuation}), (\ref{e:CurvatureScaleFluct}) and
(\ref{e:IsotropyMeasure}).  These quantities were found to be
convergent as $N_\mathrm{grid}$ increases for the evolutions on all
three manifolds: Seifert-Weber space, the lens space $L(5,2)$, and
Sixth-Turn space.  In order to reduce the number of graphs, only the
highest resolution ($N_\mathrm{graph}=28$) results for
$\llangle\,\Delta R(t)\,\rrangle$, $\llangle\,\Delta
\mathscr{R}(t)\,\rrangle$, and $\llangle\,\mathscr{I}(t)\,\rrangle$
are shown in Figs.~\ref{f:RFluctuationEv}--\ref{f:IsotrophyEv}.  These
graphs show that the homogeneity and the isotropy of these simple
cosmological models improve as the geometries expand.

An important goal of this study is to compare the simple homogeneous
and isotropic cosmological models constructed here with the standard
Friedman-Robertson-Walker (FRW) cosmological models.  The spacetime
metric for the FRW cosmological models can be written in the form,
\begin{equation}
  ds^2 = -d\eta^2 + a^2(\eta)L^2\left[d\chi^2 + \Xi^2(\chi)
    \left(d\theta^2+\sin^2\theta d\varphi^2\right)\right],
  \label{e:FriedmanMetric}
\end{equation}
where $a(\eta)$ is a dimensionless scale factor, $L$ is a
characteristic length scale, and $\Xi(\chi)$ is given by
\begin{equation}
  \Xi(\chi) = \left\{
  \begin{array}{ll}
    \sin\chi,\,\,&\mathrm{for}\,\, k=+1,\\
    \chi,\,\,&\mathrm{for}\,\, k=0,\\
    \sinh\chi,\,\,&\mathrm{for}\,\, k=-1,
  \end{array}\right.
\end{equation}
for the manifolds with positive, zero, and negative spatial curvatures
respectively.  The Einstein equations for these geometries are given
by
\begin{eqnarray}
  \left(\frac{d a}{d\eta}\right)^{\!2} &=& -\frac{k}{L^2} + \frac{\Lambda}{3}\,a^2,
  \label{e:Friedman1}
  \\
  \frac{d^{\,2}a}{d\eta^{\,2}} &=& \frac{\Lambda}{3}\,a,
  \label{e:Friedman2}
\end{eqnarray}
in a spacetime where the cosmological constant $\Lambda$ dominates
over the matter content of the universe.

There is no {\sl a priori} relationship between the $t=$ constant
hypersurfaces produced by the numerical evolutions in this study, and
the $\eta=$ constant hypersurfaces used in the descriptions of the
simple FRW cosmological models.  The $\eta=$ constant hypersurfaces
are defined by the homogeneity and isotropy of the spatial geometries
in those simple models. The evolution of the $t=$ constant
hypersurfaces is determined by the gauge conditions used to perform
these numerical evolutions. Figures~\ref{f:RFluctuationEv} and
\ref{f:QFluctuationEv} show that the $t=$ constant hypersurfaces in
the numerical evolutions have curvature scalars $R$ and $\mathscr{R}$
that are essentially constant on those surfaces.
Figure~\ref{f:IsotrophyEv} shows these geometries are also essentially
isotropic on those surfaces.  Therefore the $t=$ constant
hypersurfaces of the numerical evolutions coincide (within the
accuracy of the numerical evolutions) with the $\eta=$ constant
hypersurfaces of the simple FRW cosmological models.  In this case
$\eta$ is a function that depends only on $t$ in these models:
$\eta=f(t)$.

\begin{figure}[!t]
%  {\centering \includegraphics[height=0.32\textwidth]
%    {../Figures/ScalFactorFluctuationsEv.eps}}
  {\centering
    \includegraphics[height=0.32\textwidth]{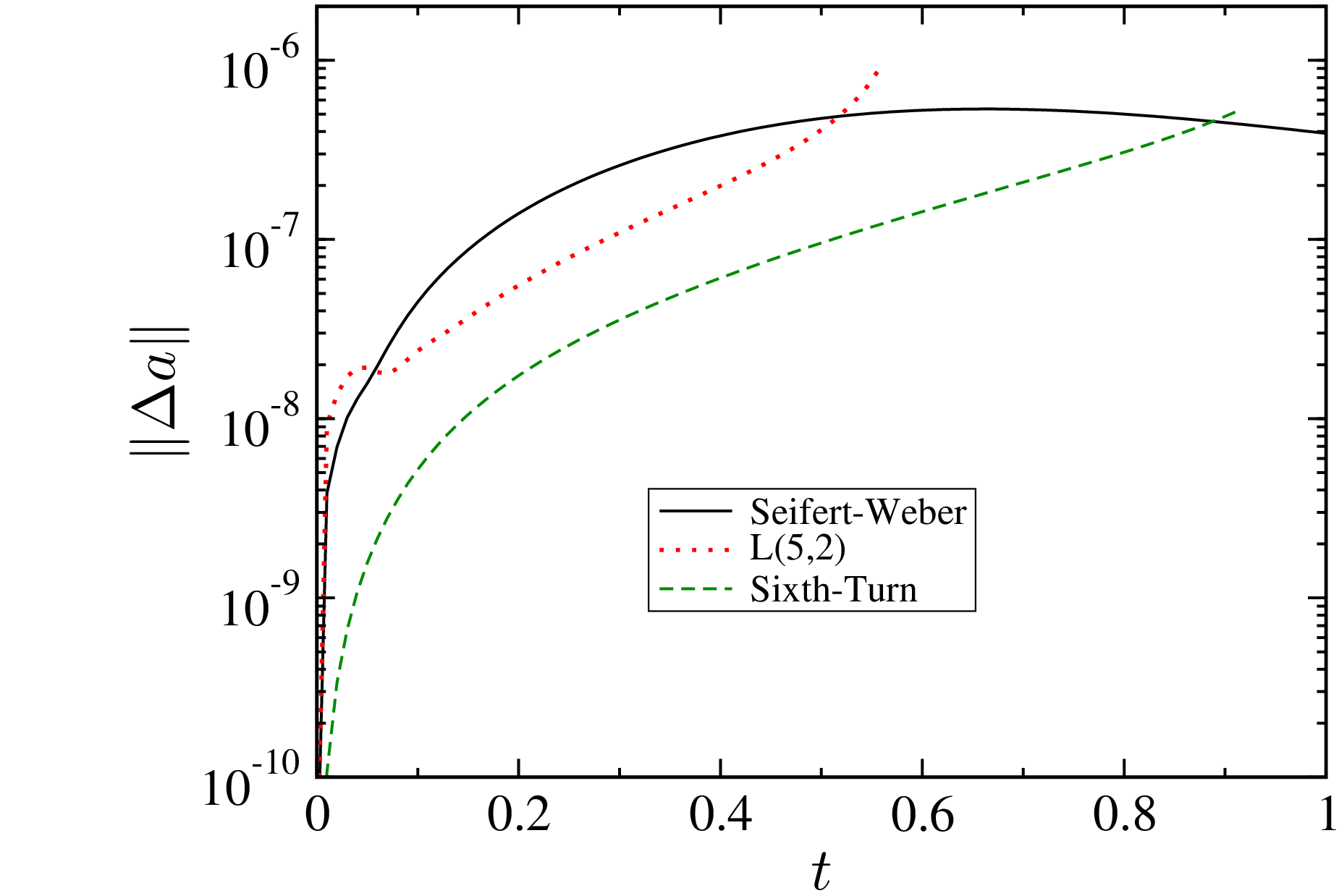}}
  { \caption{\label{f:ScalFactorFluctuationsEv} This figure
      illustrates the norm $\|\,\Delta a\,\|$ defined in
      Eq.~(\ref{e:DeltaaDef}) that measures the spatial variations in
      scale factor $a(t)$ for the $N_\mathrm{grid}=28$ solutions to
      the Einstein evolution equations on the Seifert-Weber, $L(5,2)$,
      and the Sixth-Turn manifolds.
  }}
\end{figure}

\begin{figure}[!t]
%   {\centering \includegraphics[height=0.32\textwidth]
%     {../Figures/LapseFluctuationEv.eps}}
   {\centering \includegraphics[height=0.32\textwidth]
        {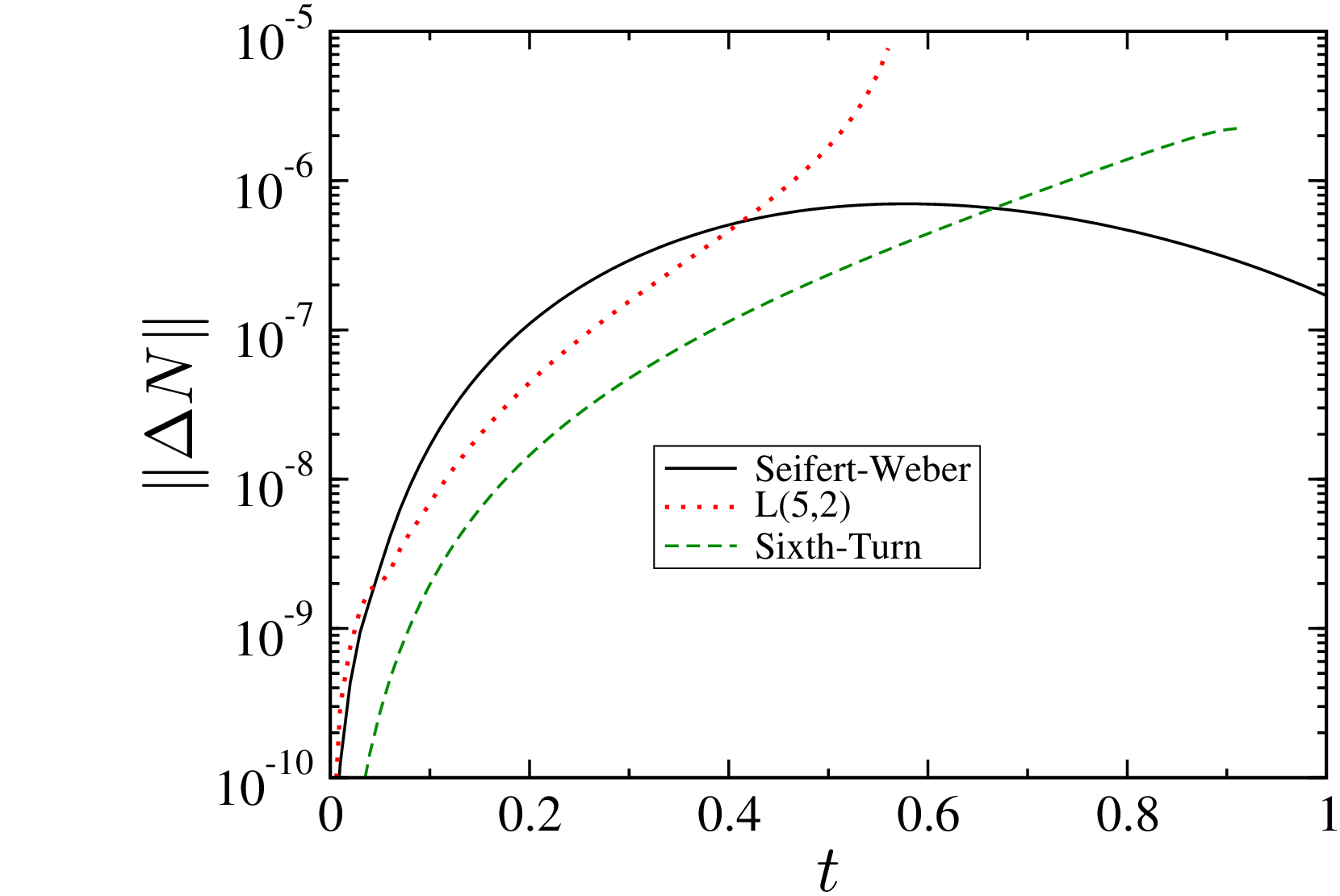}}
  { \caption{\label{f:LapseFluctuationEv} This figure illustrates the
      norm $\|\,\Delta N\,\|$ defined in Eq.~(\ref{e:DeltaNDef}) that
      measures the spatial variations in lapse $N(t)$ for the
      $N_\mathrm{grid}=28$ solutions to the Einstein evolution
      equations on the Seifert-Weber, $L(5,2)$, and the Sixth-Turn
      manifolds.  }}
\end{figure}

\begin{figure}[!t]
%  {\centering \includegraphics[height=0.32\textwidth]
%    {../Figures/SW-ScalFactorFluctuationsEv.eps}}
  {\centering
    \includegraphics[height=0.32\textwidth]{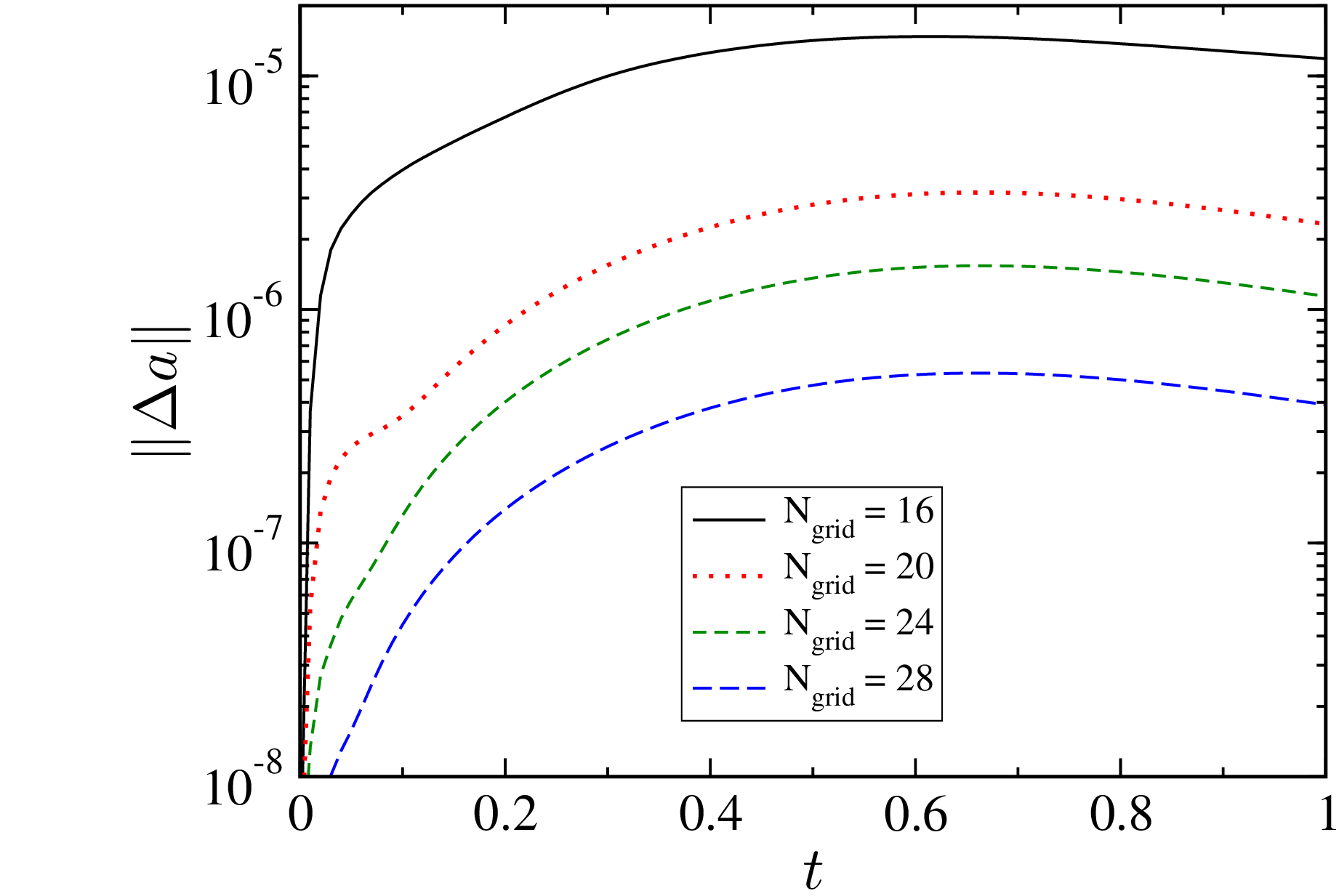}}
  { \caption{\label{f:SW-ScalFactorFluctuationsEv} This figure
      illustrates the numerical convergence of the norm $\|\,\Delta
      a\,\|$ defined in Eq.~(\ref{e:DeltaaDef}) as the spatial
      resolution $N_\mathrm{grid}$ is increased for solutions to the
      Einstein evolution equations on the Seifert-Weber manifold.  }}
\end{figure}

\begin{figure}[!t]
%   {\centering \includegraphics[height=0.32\textwidth]
%     {../Figures/E5-LapseFluctuationEV.eps}}
   {\centering \includegraphics[height=0.32\textwidth]
        {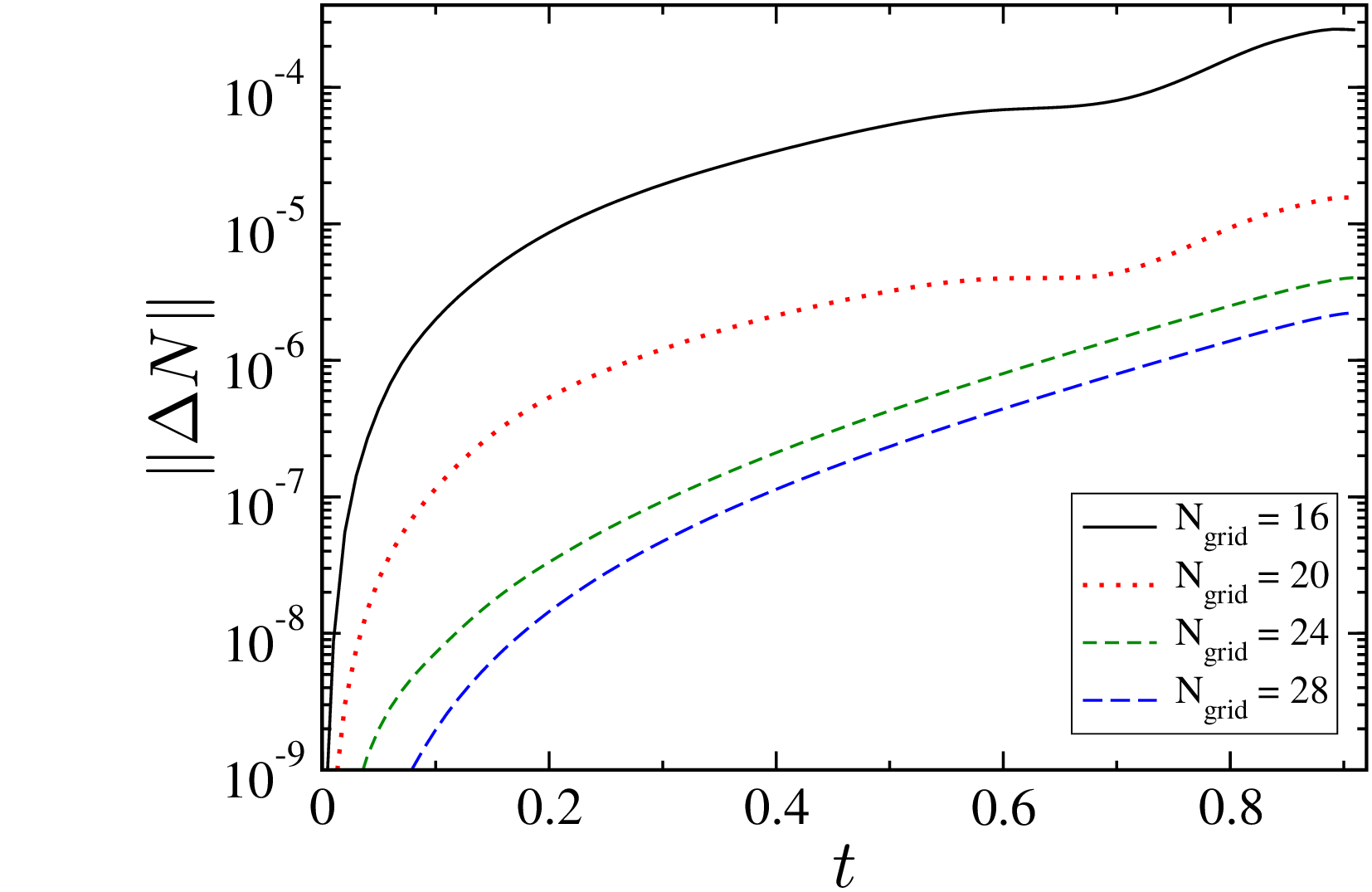}}
  { \caption{\label{f:E5-LapseFluctuationEv} This figure
        illustrates the numerical convergence of the norm $\|\,\Delta
        N\,\|$ defined in Eq.~(\ref{e:DeltaNDef}) as the spatial
        resolution $N_\mathrm{grid}$ is increased for solutions to the
        Einstein evolution equations on the Sixth-Turn manifold.  }}
\end{figure}

To facilitate comparisons between the numerical cosmological solutions
computed in this study and the simple analytical FRW cosmological
models it will be helpful to define a scale factor $a(t)$ for the
numerical models that measures how the determinant of the spatial
metric, $g(t)=\det g_{ij}(t)$, evolves with time:
\begin{equation}
  a^6(t)=\frac{g(t)}{g(0)}.
  \label{e:ScaleFactorDef}
\end{equation}
The spatial average $\llangle a(t)\rrangle$, defined by
\begin{equation}
  \llangle a(t)\rrangle^3 = \frac{\int\sqrt{g(t)}\,d^{\,3}x}
          {\int\sqrt{g(0)}\,d^{\,3}x}
          = \frac{\mathcal{V}(t)}{\mathcal{V}(0)},
  \label{e:ScaleFactorAverage}
\end{equation}
can therefore serve as a generalization of the scale factor used in
the analytic FRW cosmological models.  The spatial variations in
$a(t)$ can be measured with respect to its spatial average, $\llangle
a(t)\rrangle$, by evaluating the norm $\|\,\Delta a\,\|$ defined by
\begin{equation}
  \|\,\Delta a\,\|^2 = \frac{\int \left[a(t)-\llangle a(t)\rrangle\right]^2
  \sqrt{g}\,d^{\,3}x}{\llangle a(t)\rrangle^2\,\mathcal{V}(t)}.
  \label{e:DeltaaDef}
\end{equation}
Similarly the homogeneity of the lapse $N$ can be evaluated:
\begin{eqnarray}
  \|\,\Delta N\,\|^2 &=& \frac{\int \left[N(t) - \llangle N(t)\rrangle\right]^2
  \sqrt{g}\,d^{\,3}x}{\llangle N(t)\rrangle^2\,\mathcal{V}(t)},
  \label{e:DeltaNDef}
\end{eqnarray}
where $\llangle N(t)\rrangle$ is defined by,
\begin{eqnarray}
\llangle N(t) \rrangle &=& \frac{\int N(t)   
  \sqrt{g}\,d^{\,3}x}{\mathcal{V}(t)}.
  \label{e:NAverageDef}
\end{eqnarray}
Figures~\ref{f:ScalFactorFluctuationsEv} and
\ref{f:LapseFluctuationEv} demonstrate that the scale factor $a(t)$
and the lapse $N(t)$ are essentially constants on each $t=$ constant
hypersurface in these numerical evolutions. The spatial variations in
the scale factor and the lapse converge toward zero as the spatial
resolution $N_\mathrm{grid}$ increases in all these
evolutions. Figures~\ref{f:SW-ScalFactorFluctuationsEv} and
\ref{f:E5-LapseFluctuationEv} illustrate these convergence results for
the Seifert-Weber and the Sixth-Turn manifolds.

\begin{figure}[!t]
%  {\centering \includegraphics[height=0.32\textwidth]
%    {../Figures/FriedmanFit.eps}}
  {\centering \includegraphics[height=0.32\textwidth]
    {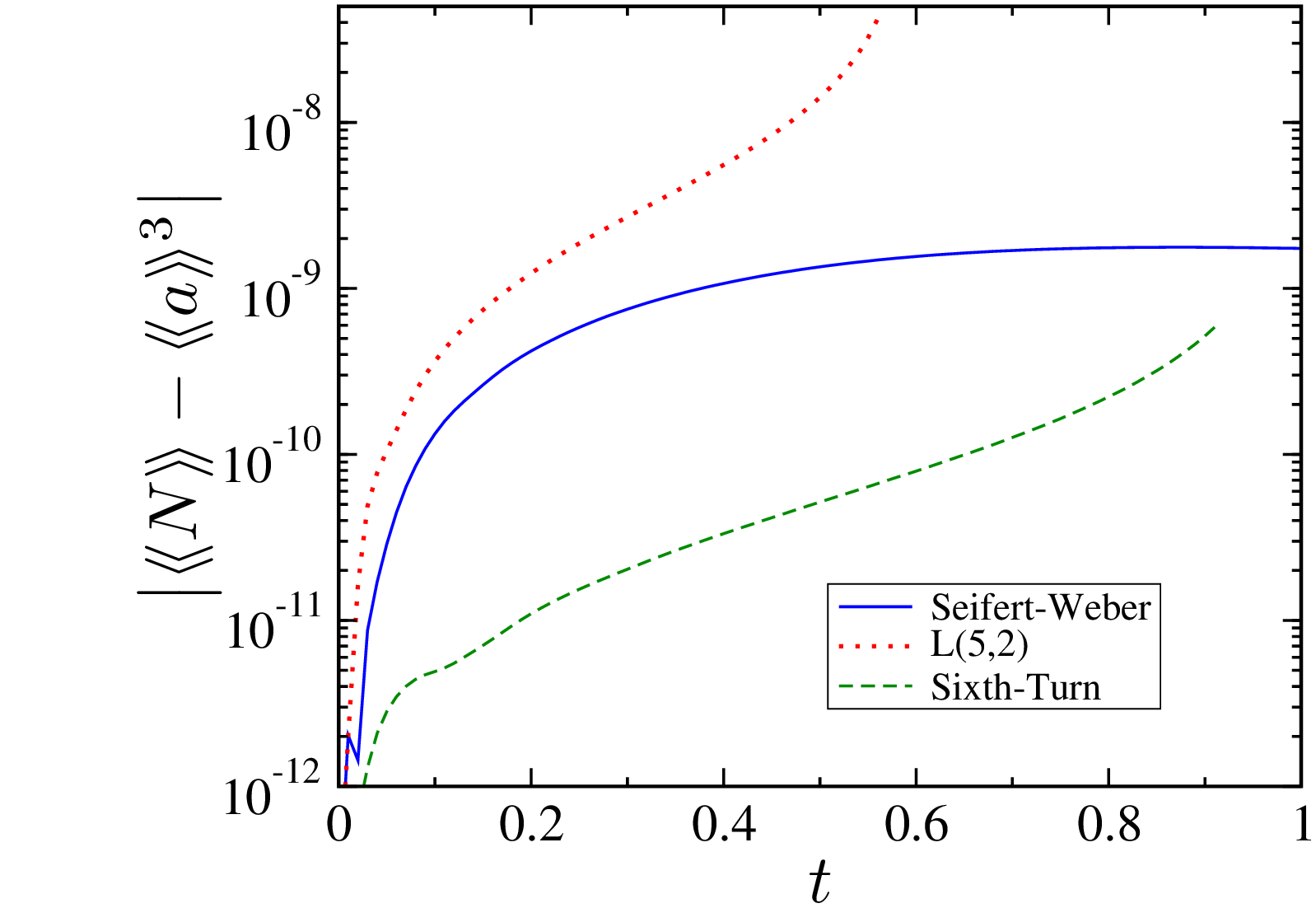}}
  { \caption{\label{f:FriedmanFit} This figure illustrates the
      difference between the average value of the Lapse $\llangle
      N(t)\rrangle$ defined in Eq.~(\ref{e:ADMMetric}) and the average
      scale factor $\llangle a(t)\rrangle^3$ defined in
      Eq.~(\ref{e:ScaleFactorAverage}) for the $N_\mathrm{grid} = 28$
      solution to the Einstein evolution equations on the Seifert-Weber,
      $L(5,2)$, and the Sixth-Turn manifolds.}}
\end{figure}

\begin{figure}[!t]
%  {\centering \includegraphics[height=0.32\textwidth]
%    {../Figures/L52-FriedmanFit.eps}}
    {\centering \includegraphics[height=0.32\textwidth]
    {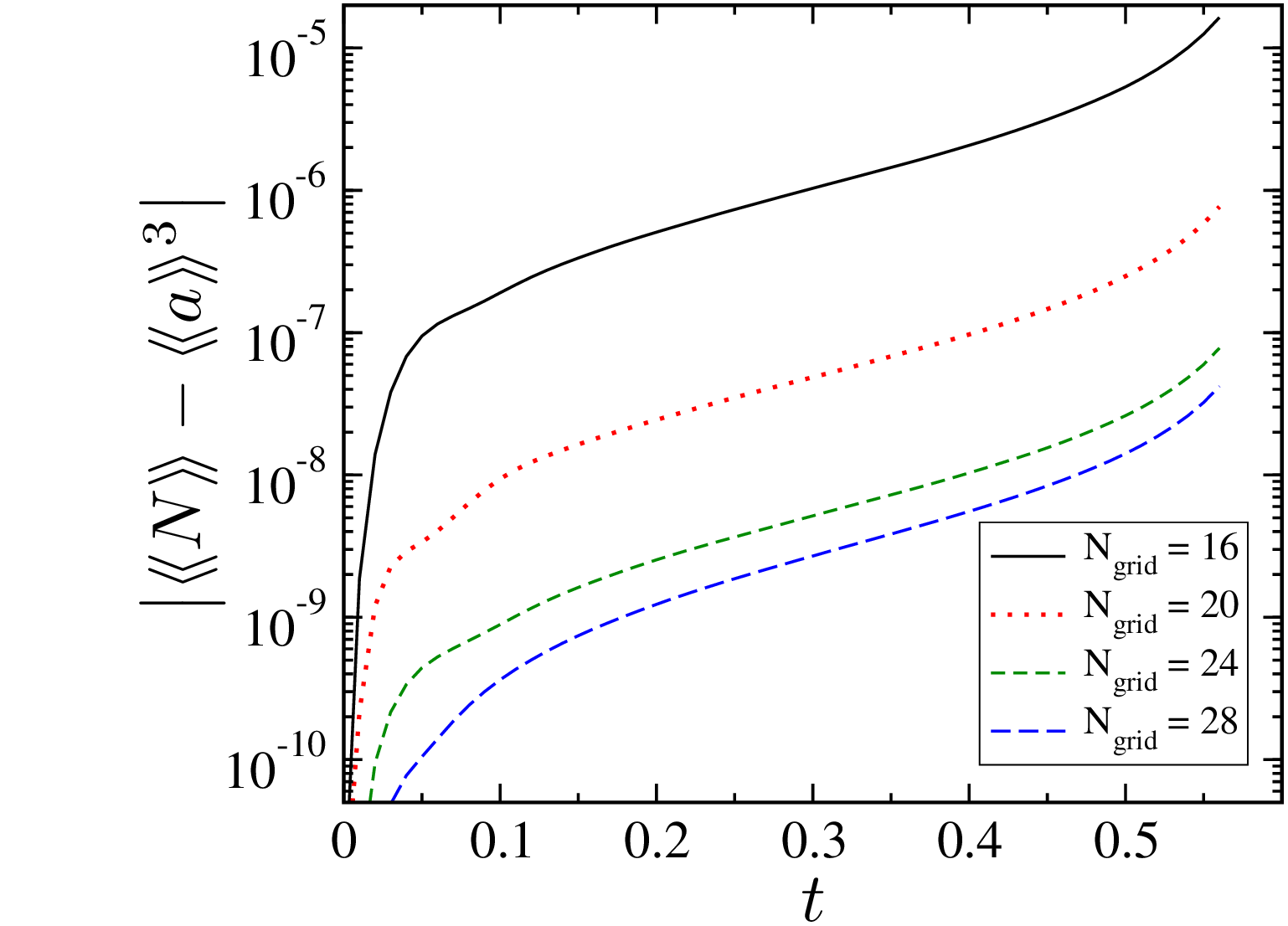}}
%  {\centering
%    \includegraphics[height=0.32\textwidth] {./Fig16.eps}}
  { \caption{\label{f:L52-FriedmanFit} This figure
        illustrates the numerical convergence of the difference
        between the average value of the Lapse $\llangle N(t)\rrangle$
        defined in Eq.~(\ref{e:ADMMetric}) and the average scale
        factor $\llangle a(t)\rrangle^3$ defined in
        Eq.~(\ref{e:ScaleFactorAverage}) as the numerical resolution
        $N_\mathrm{grid}$ is increased for solutions to the Einstein
        evolution equations on the $L(5,2)$ manifold.}}
\end{figure}

It is interesting to compare the numerical evolution of $\llangle
a(t)\rrangle$ with the analogous solutions for $a(\eta)$ from
Eqs.~(\ref{e:Friedman1}) and (\ref{e:Friedman2}).  These comparisons
are complicated by the fact that the gauge used in the numerical
evolutions does not keep the lapse $N(t)$ fixed at its initial value.
Consequently the relationship between the numerical time coordinate,
$t$, and the time coordinate, $\eta$, used on the FRW cosmological
models must be determined before comparisons between the cosmological
models can be made.  Figure~\ref{f:FriedmanFit} shows that the gauge
condition that determines that the evolution of the lapse results in
the condition $\llangle N(t)\rrangle \approx \llangle a(t)\rrangle^3$
in these evolutions.  Figure~\ref{f:L52-FriedmanFit} illustrates the
numerical convergence of this result for the $L(5,2)$ manifold as the
spatial resolution $N_\mathrm{grid}$ increases.  Comparing
Eq.~(\ref{e:ADMMetric}) with Eq.~(\ref{e:FriedmanMetric}) shows that
$d\eta/dt = N(t)$.  Therefore $d\eta/dt=a^3(t)$, so the FRW version of
the Einstein evolution Eqs.~(\ref{e:Friedman1}) and
(\ref{e:Friedman2}) using the time coordinate $t$ used in this study
can be written in the forms,
\begin{eqnarray}
  \left(\frac{da_\mathrm{\scriptscriptstyle FRW}}{dt}\right)^2
  &=& a^6_\mathrm{\scriptscriptstyle FRW}\left(-\frac{k}{L^2} +
  \frac{\Lambda}{3}a^2_\mathrm{\scriptscriptstyle FRW}\right),
  \label{e:FriedmanEq2}\\
  \frac{d^2a_\mathrm{\scriptscriptstyle FRW}}{dt^2}
  &=& \frac{3}{a_\mathrm{\scriptscriptstyle FRW}}
  \left(\frac{da_\mathrm{\scriptscriptstyle FRW}}{dt}\right)^2 +
  \frac{\Lambda}{3}a^7_\mathrm{\scriptscriptstyle FRW}.
  \label{e:FriedmanEq2.5}
\end{eqnarray}

\begin{figure}[!t]
%  {\centering \includegraphics[height=0.32\textwidth]
%    {../Figures/FRW-Fits.eps}}
  {\centering \includegraphics[height=0.32\textwidth]{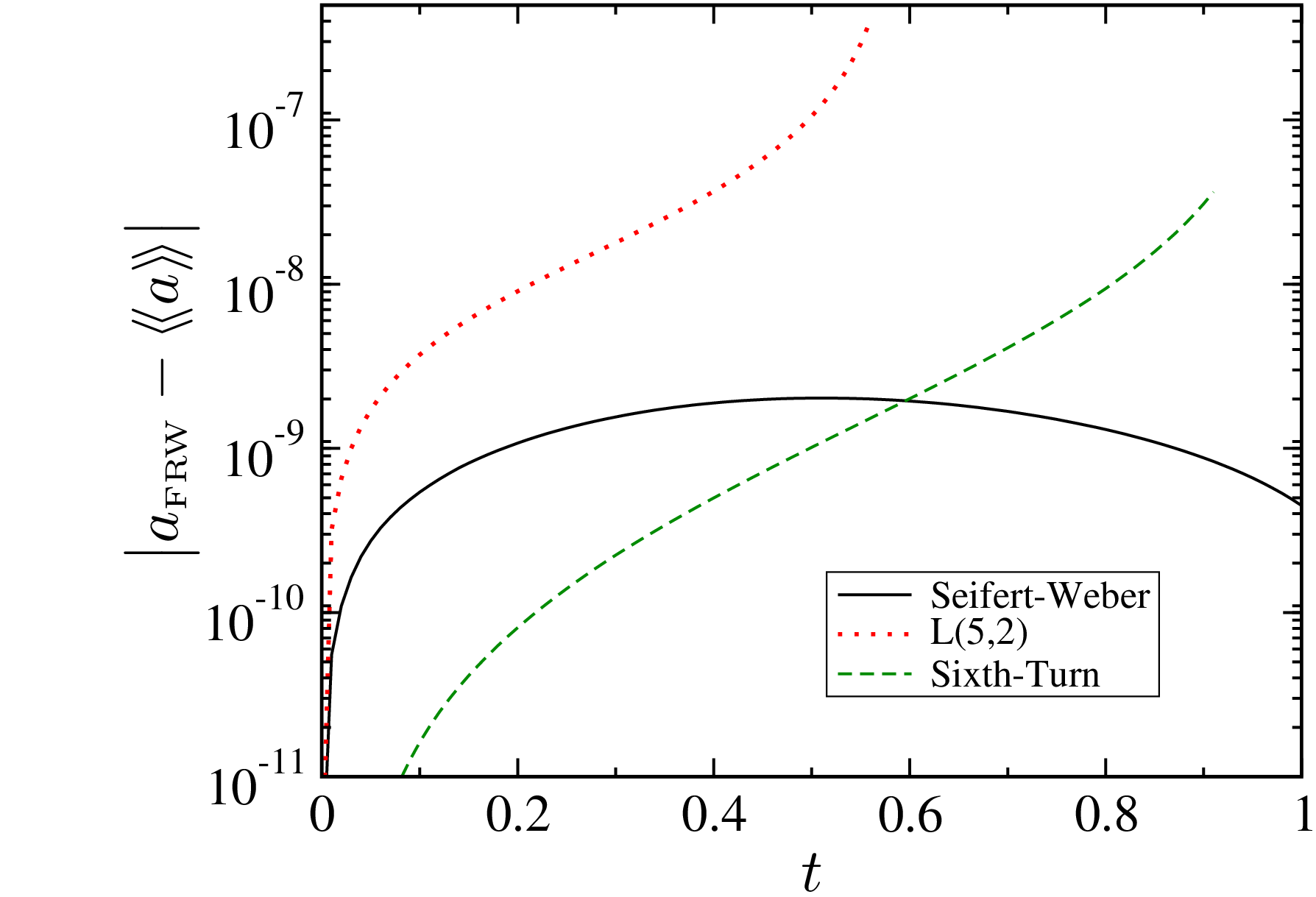}}
  { \caption{\label{f:FRW-Fits} This figure illustrates the
      difference between the FRW scale factor
      $a_\mathrm{\scriptscriptstyle FRW}(t)$ determined by solving
      Eq.~(\ref{e:FriedmanEq2.5}) and the average scale factor
      $\llangle a(t)\rrangle$ from the $N_\mathrm{grid} = 28$
      solutions to the Einstein evolution equations on the Seifert-Weber,
      $L(5,2)$ and the Sixth-Turn manifolds.  }}
\end{figure}

The solutions, $a_{\scriptscriptstyle \mathrm{FRW}}(t)$, to these FRW
cosmology equations can now be compared directly with the $\llangle
a(t)\rrangle$ for the simple cosmological models computed numerically
for this study.  In the $k=0$ case Eq.~(\ref{e:FriedmanEq2})
simplifies to $da_\mathrm{\scriptscriptstyle FRW}/dt =
a^4_\mathrm{\scriptscriptstyle FRW}\sqrt{\Lambda/3}$, which has the
analytic solution $a_\mathrm{\scriptscriptstyle
  FRW}(t)=(1-\sqrt{3\Lambda}\,t)^{-1/3}$. (This result was also found
in Ref.~\cite{Zhang2026a} for a numerical cosmological model obtained
with the same evolution gauge condition on the non-orientable manifold
with spatial topology $P^2\#P^2\times S^1$.)  In the $k=\pm 1$ cases
Eq.~(\ref{e:FriedmanEq2}) does not have simple analytical solutions.
In those cases it is more convenient to solve the second-order
Eq.~(\ref{e:FriedmanEq2.5}) numerically.  The initial values of
$a_\mathrm{\scriptscriptstyle FRW}$ and $da_\mathrm{\scriptscriptstyle
  FRW}/dt$ at $t=0$ needed to solve Eq.~(\ref{e:FriedmanEq2.5}) are
fixed by the values of $\llangle a(0) \rrangle$ and $d\llangle a(0)
\rrangle/dt$ from the numerical evolutions at that time.  The
comparisons between the $a_\mathrm{\scriptscriptstyle FRW}(t)$
computed in this way and $\llangle a(t)\rrangle $ are illustrated in
Fig.~\ref{f:FRW-Fits} for the three different exemplar manifolds
included in this study.  These results show that the simple
cosmological models computed for this study have the same expansion
histories as the standard FRW models within the accuracies of the
numerical models.
%
%%%%%%%%%%%%%%%%%%%%%%%%%%%%%%%%%%%%%%%%%%%%%%%%%%%%%%%%%%%%%%%%%%%%%%%%%%%%%%%
\section{Discussion}
\label{s:Discussion}
%%%%%%%%%%%%%%%%%%%%%%%%%%%%%%%%%%%%%%%%%%%%%%%%%%%%%%%%%%%%%%%%%%%%%%%%%%%%%%%

The motivation for this study was to determine whether Ricci flow
could be useful in the construction of initial data for realistic
cosmological models on manifolds with non-trivial spatial topologies.
The results presented here show that convergent numerical solutions to
the Ricci flow equation representing homogeneous and isotropic spatial
geometries can be constructed successfully on manifolds with a range
of topologies.  The results of these Ricci flow evolutions were used
in this study to prepare initial data for the Einstein evolution
equations, which were then solved numerically to construct simple
cosmological models.  The resulting models were shown to be
homogeneous and isotropic, and to have the same expansion histories as
the analogous Friedman-Robertson-Walker cosmologies.  These results
demonstrate the usefulness of Ricci flow in the construction of
initial data for realistic cosmological models.  The agreement
demonstrated here between the complicated numerical results prepared
for this study and the analogous results for simple
Friedman-Robertson-Walker cosmological models shows the robustness of
our numerical methods and their implementation in the \verb!SpEC!
code.

The Ricci flow equation is very inefficient to solve numerically
because severe limitations on the time step size are required to
ensure stable numerical evolutions, see Ref.~\cite{Canuto1988,
  Boyd2000}.  A similar problem exists with the numerical solution of
elliptic systems of equations like the Einstein initial value equations.
In some cases these inefficiency problems have been overcome for the
constraint equations by transforming those elliptic systems into
damped hyperbolic equations.  This hyperbolic relaxation approach has
been used successfully in some cases to significantly improve the
efficiency of producing numerical solutions to the Einstein
constraints, see Refs.~\cite{Ruter2018, Zhang2024}.  It would be
interesting to see whether the parabolic Ricci flow equation could be
converted into a damped hyperbolic equation that significantly
improves computational efficiency in this case as well.

The harmonic gauge used to produce the numerical evolutions for this
study produced cosmological models with $N(t)=a^3(t)$.  This results
in evolutions in which the scale factor, and therefore the spatial
volumes, approaches infinity at finite values of the time parameter.
For example in the Sixth-Turn space case $a_\mathrm{FRW}(t)=
(1-\sqrt{3}\Lambda)^{-1/3}$.  This fact severely limits our ability to
evolve models numerically long enough to allow the scale factors to
increase to the very large values that have been observed in our
universe.  Therefore it would be very useful to try to develop a gauge
condition better suited for these simple cosmological models.

%%%%%%%%%%%%%%%%%%%%%%%%%%%%%%%%%%%%%%%%%%%%%%%%%%%%%%%%%%%%%%%%%%%%%%%%%%%%%%%

\appendix
%%%%%%%%%%%%%%%%%%%%%%%%%%%%%%%%%%%%%%%%%%%%%%%%%%%%%%%%%%%%%%%%%%%%%%%%%%%%%%%
\appendix
%%%%%%%%%%%%%%%%%%%%%%%%%%%%%%%%%%%%%%%%%%%%%%%%%%%%%%%%%%%%%%%%%%%%%%%%%%%%%%%
\section{Covariant First-Order Symmetric-Hyperbolic Einstein System}
\label{s:SymmetricHyperbolicSystem}
%%%%%%%%%%%%%%%%%%%%%%%%%%%%%%%%%%%%%%%%%%%%%%%%%%%%%%%%%%%%%%%%%%%%%%%%%%%%%%%
\renewcommand{\theequation}{A.\arabic{equation}} % Redefines equation numbering
\setcounter{equation}{0} % Resets the equation counter for the appendix

This study uses a covariant representation of the Einstein equations
that enforces generalized harmonic gauge conditions, see
Ref.~\cite{Lindblom2014}.  The covariance of this representation is
essential for the study of solutions on manifolds with multiple
multi-cube coordinate patches, such as the manifolds studied here.
All the dynamical fields in this representation are tensors, so they
can be transformed in a straightforward way across the interfaces
between coordinate patches.  The dynamical fields, $u^\alpha$, for the
first-order symmetric-hyperbolic representation of the Einstein
equations used in this study are the collection of tensor fields,
\begin{equation}
  u^\alpha = \left\{ \psi_{ab}, \Pi_{ab}, \Phi_{iab}\right\},
  \label{e:UalphaDef}
\end{equation}
were $\psi_{ab}$ is the spacetime metric, and the tensors $\Pi_{ab}$
and $\Phi_{iab}$ represent its first derivatives,
\begin{eqnarray}
  \Pi_{ab} &=& -t^c\tilde\nabla_c\psi_{ab},
  \label{e:PiDef}\\
  \Phi_{iab}&=& \tilde\nabla_i\psi_{ab}. 
  \label{e:PhiDef}
\end{eqnarray} 
The derivative $\tilde\nabla_c$ used in these expressions is the
torsion-free covariant derivative compatible with the reference metric
$\tilde\psi_{ab}$: $\tilde\nabla_c\tilde\psi_{ab}=0$.  This reference
metric is also used to define the Jacobians used to transform tensor
fields across the boundary interfaces between multi-cube regions.  The
vector $t^c$ represents the unit timelike normal ($t^a t^b
\psi_{ab}=-1$) to the spacelike hypersurfaces used to perform the
evolutions.  The lower range of Latin indices, e.g. $a$, $b$, $c$,
..., range over the four spacetime coordinates, while the higher range
of indices, e.g. $i$, $j$, $k$, ..., range over the three spatial
coordinates on each $t=$ constant hypersurface.  The Greek indices,
e.g. $\alpha$, $\beta$, $\gamma$, range over the fifty components of
the dynamical tensor fields $u^\alpha$.  The Einstein equations for
these fields can be written in the first-order form
\begin{equation}
  \partial_t u^\alpha
  + A^{k\alpha}{}_\beta({\bf u})\tilde\nabla_ku^\beta
  =F^\alpha({\bf u},{\bf \tilde\Gamma}, {\bf \partial\tilde\Gamma}),
  \label{e:EvolutionEqs}
\end{equation}
where the tensors $ A^{k\alpha}{}_\beta({\bf u})$ and $F^\alpha({\bf
  u},{\bf \tilde\Gamma}, {\bf \partial\tilde\Gamma})$ depend in
complicated ways on the dynamical fields $u^\alpha$, the reference
connection $\tilde\Gamma^c{}_{ab}$ and its derivatives
$\partial_d\tilde\Gamma^c{}_{ab}$.  The detailed expressions for this
representation of the covariant Einstein evolution system are given in
Ref.~\cite{Lindblom2014}.

The spacetime coordinates are determined in the generalized harmonic
representations of Einstein's equation by imposing a condition on the
difference between the connection $\tilde\Gamma^c{}_{ab}$ associated
with the reference metric $\tilde \psi_{ab}$ and the physical
connection $\Gamma^c{}_{ab}$ associated with the physical spacetime
metric $\psi_{ab}$.  In particular the difference between these
connections is fixed by a gauge source function $H_a$:
\begin{equation}
H_a =  -\psi_{ad}\,\psi^{bc}\left(\Gamma^d{}_{bc}-\tilde\Gamma^d{}_{bc}\right).
  \label{e:HarmonicGageDef}
\end{equation}
(Note that the difference between any two connections is a tensor.)
In general $H_a$ may be any function that depends on the physical
metric $\psi_{ab}$ (but not its derivatives) and the reference metric
$\tilde\psi_{ab}$.  The gauge condition used for the numerical
evolutions performed in this study is the simple harmonic condition,
$H_a=0$.

The generalized harmonic evolution system contains a number of
constraints.  In particular the gauge condition,
Eq.~(\ref{e:HarmonicGageDef}) is in effect a constraint
\begin{equation}
  \mathcal{C}_a =H_a +
  \psi_{ad}\psi^{bc}\left(\Gamma^d{}_{bc}-\tilde\Gamma^d{}_{bc}\right)
  \label{e:C1Def}
\end{equation}
on the dynamical fields.  In addition, the equation that defines
$\Phi_{iab}$, Eq. (\ref{e:PhiDef}), is also a constraint
\begin{equation}
  \mathcal{C}_{iab} = \tilde\nabla_i\psi_{ab}-\Phi_{iab}.
  \label{e:C3Def}
\end{equation}
These primary constraints, $\mathcal{C}_a$ and $\mathcal{C}_{iab}$,
satisfy a second-order system of evolution equations as a consequence
of the Einstein evolution system, Eq.~(\ref{e:EvolutionEqs}). (See
Ref.~\cite{Lindblom2014} for details.)  This second-order constraint
evolution system can be converted to a first-order symmetric
hyperbolic system by introducing the secondary constraints,
\begin{eqnarray}
  \mathcal{F}_a &=& t^c\nabla_c\mathcal{C}_a,
  \label{e:F1Def}\\
  \mathcal{C}_{ia} &=& \nabla_i\mathcal{C}_a,
  \label{e:C2Def}\\
  \mathcal{C}_{ijab}&=&2\tilde\nabla_{[i}\mathcal{C}_{j]ab}.
  \label{e:C4Def}
\end{eqnarray}
The symmetric-hyperbolic evolution equation for the collection of
constraints,
\begin{equation}
  \mathcal{C}^\alpha=\left\{\mathcal{C}_a,\mathcal{C}_{iab},
  \mathcal{F}_a,\mathcal{C}_{ia},\mathcal{C}_{ijab}\right\},
  \label{e:ConstraintTotal}
\end{equation}
ensures that solutions to the Einstein Eq.~(\ref{e:EvolutionEqs}) that
satisfy the constraints, $\mathcal{C}^\alpha$, on an initial surface
will satisfy them throughout the evolution of those initial data.
Numerical solutions to the equations will of course contain (hopefully
small) violations of these constraints.  This study monitors the size
of these constraint violations by evaluating a norm constructed from
the composite constraint, $\mathcal{C}_\psi$, defined by,
\begin{eqnarray}
  \mathcal{C}_\psi^2 &=& \delta^{ab}\left(\mathcal{C}_a\mathcal{C}_b
  +\mathcal{F}_a\mathcal{F}_b\right)\nonumber\\
&&  +\tilde g^{ij}\delta^{ab}\delta^{cd}\left(\mathcal{C}_{iac}\mathcal{C}_{jbd}
    +\tfrac{1}{4}\tilde g^{kl}\mathcal{C}_{ikac}\mathcal{C}_{jlbd}\right).
    \qquad
  \label{e:CpsiDef}
\end{eqnarray}
This study evaluates a norm of $\mathcal{C}_\psi$,
$\|\,\mathcal{C}_\psi\,\|$, defined by
\begin{equation}
  \|\,\mathcal{C}_\psi\,\|^2 = \frac{\int \mathcal{C}_\psi^2\sqrt{g}\,d^{\,3}x}
  {\int\mathcal{N}_\psi^2\sqrt{g}\,d^{\,3}x}.
  \label{e:CpsiNormDef}
\end{equation}
The quantity that appears in the denominator, $\mathcal{N}_\psi$,
is defined by
\begin{eqnarray}
  \mathcal{N}_\psi^2&=& \delta^{ab}\delta^{cd}\tilde g^{ij}\left(
  \partial_i\psi_{ac}\partial_j\psi_{bd}
  +  \partial_i\Pi_{ac}\partial_j\Pi_{bd}\right)\nonumber\\
  &&+\delta^{ab}\delta^{cd}\tilde g^{ij}
  \tilde g^{kl}\partial_i\Phi_{kac}\partial_j\Phi_{lbd}.
  \label{e:NpsiDef}
\end{eqnarray}
It has been included in the definition of $\|\,\mathcal{C}_\psi\,\|$ to 
provide a dimensionless measure of the fractional errors due to
constraint violations in the numerical solutions to the Einstein
evolution system.

%%%%%%%%%%%%%%%%%%%%%%%%%%%%%%%%%%%%%%%%%%%%%%%%%%%%%%%%%%%%%%%%%%%%%%%%%%%%%%%
\begin{acknowledgements}
  We thank Michael Holst for providing the computational resources
  used to perform the numerical calculations reported in this study.
  L.L. was supported in part by grant No. 2407545 from the National
  Science Foundation to the University of California at San Diego,
  USA. F.Z. was supported by the National Key Research and Development
  Program of China grant 2023YFC2205801, and the National Natural
  Science Foundation of China grants 12433001 and 12021003.
\end{acknowledgements}
%%%%%%%%%%%%%%%%%%%%%%%%%%%%%%%%%%%%%%%%%%%%%%%%%%%%%%%%%%%%%%%%%%%%%%%%%%%%%%%

%%%%%%%%%%%%%%%%%%%%%%%%%%%%%%%%%%%%%%%%%%%%%%%%%%%%%%%%%%%%%%%%%%%%%%%%%%%%%%%
% References
%%%%%%%%%%%%%%%%%%%%%%%%%%%%%%%%%%%%%%%%%%%%%%%%%%%%%%%%%%%%%%%%%%%%%%%%%%%%%%%

\bibliography{../References/References}

%%%%%%%%%%%%%%%%%%%%%%%%%%%%%%%%%%%%%%%%%%%%%%%%%%%%%%%%%%%%%%%%%%%%%%%%%%%%%%%
\end{document}